# Dynamic competition between phason and amplitudon observed by ultrafast multimodal scanning tunneling microscopy

Seokjin Bae[1†], Arjun Raghavan[1†], Soyeun Kim[1,2], Kejian Qu[1], Chengxi Zhao,[3], Daniel P. Shoemaker[3], Ziqiang Wang[4], Fahad Mahmood[1], Barry Bradlyn[1], Vidya Madhavan,[1*]

[1] Department of Physics, Grainger College of Engineering, University of Illinois Urbana-Champaign, Urbana, IL 61801, USA.

[2] Department of Physics and Chemistry, Daegu Gyeongbuk Institute of Science and Technology (DGIST), Daegu 42988, South Korea

[3] Department of Materials Science and Engineering, Grainger College of Engineering, University of Illinois Urbana-Champaign, Urbana, IL 61801, USA

[4] Department of Physics, Boston College, Chestnut Hill, Massachusetts 02467, USA

The intertwining between two ordered states that arise from the same interactions is reflected in the dynamics of their coupled collective excitations. While the equilibrium phase diagram resulting from such intertwined orders has been extensively studied, the dynamic competition between non-equilibrium modes is a largely unexplored territory. Here, we introduce a multimodal STM-based pump-probe technique, that combines ultrafast tunneling microscopy (USTM), ultrafast point-contact spectroscopy (UPC), and optical pump-probe reflectance (OPPR) on femtosecond timescale, all within a single instrument. Using this platform, we investigate the collective excitations of the unconventional charge density wave insulator $(TaSe_4)_2I$. Our UPC measurements reveal charge oscillations at 0.22 THz, with a temperature dependence that matches the theoretically predicted behavior of the long-sought massive phason gaining mass through the Anderson-Higgs mechanism. Unexpectedly, the data also reveals a second mode at 0.11 THz exhibiting similar temperature and polarization dependence with comparable mode intensity. These features, along with the robust 1/2 frequency ratio locking suggest that the 0.11 THz phason is a 'daughter mode' that arises from the splitting of the 0.22 THz massive phason into two massless phasons via parametric amplification, analogous to the decay of a neutral pion into two photons. Strikingly, comparison with OPPR data reveals that the daughter phason competes with and suppresses the amplitudon at proximate frequency. Our studies reveal an unexplored mechanism for the generation and extinction of collective excitations in quantum materials and pave the way for a microscopic understanding of ultrafast phenomena.

†These authors contributed equally to this work

*Correspondence to: vm1@illinois.edu

A phason is one of the two fundamental collective excitations of an incommensurate charge density wave (CDW) state, with the other being an amplitudon (Fig. 1a)[1]. It represents a sliding motion of the charge modulation. In typical CDW systems, the phason is a massless excitation with a gapless spectrum or a very narrow gap (on the order of a few tens of GHz) generated by the pinning potential around defects[2,3]. However, as proposed in a series of papers beginning with the seminal work of Lee and Fukuyama[4], in insulators at low temperatures, where long-range Coulomb interactions are poorly screened due to the lack of thermally excited quasiparticles, it is theorized that the phason becomes massive and gapped at optical frequencies[5,6] (Fig. 1b). This phenomenon is analogous to the Anderson-Higgs mechanism[7–10], where photons acquire an effective mass when coupled to the Goldstone mode of a superconductor, which is also the predicted fate of a phason in the presence of strong Coulomb interactions. To test this prediction directly, two key experimental requirements must be satisfied. First, to prevent the screening of long-range Coulomb interactions, the material must be an insulator with minimal quasiparticle excitations at the relevant temperatures. This is a challenging requirement, as the formation of a CDW order typically leads to only a partial gapping of the Fermi surface, resulting in a metallic state[11–14]. Second, as the phase of a CDW oscillates, the charge density also oscillates, inducing an ultrafast current (Fig. 1c). Therefore, to directly measure the massive phasons, one would need to capture local current oscillations within a CDW domain[15] with femtosecond time resolution.

To investigate the existence and characteristics of massive phasons and their interactions with other modes, we introduce a multimodal ultrafast STM which enables ultrafast tunneling, ultrafast point-contact (UPC), and optical pump-probe reflectivity (OPPR) with femtosecond resolution. The technique relies on measuring pair-pulse correlations[16–18] of femtosecond near-infrared pulsed laser excitations and builds upon recent advancements in THz and infrared STM[19–27]. In this study, we apply these techniques to the unconventional CDW insulator $(TaSe_4)_2I$, which has recently been shown to emit radiation at 0.22 THz[28], thought to originate from a massive phason.

$(TaSe_4)_2I$ is a quasi-1D crystal consisting of Ta-chains along the c-axis which are encapsulated by Se-rectangles and spaced by iodine atoms (Fig. 1d). Lattice constants are a = b = 9.54 Å and c = 12.77 Å[29]. The cleavage plane is the (110) plane which reveals the iodine surface. Nearly half of the iodine atoms on the cleaving plane are taken away during cleaving process[15] (Supplementary Fig. 1), the remaining iodine atoms form a square lattice which is ~ 45 degree rotated from the lattice vector of the original (conventional) unit cell as seen in the topographic image at 77 K and its Fourier transform (FT) in Fig. 1e. An incommensurate CDW with the wavelength of ≈ 9.1 nm is visible in the topography which forms below the CDW transition temperature $T_c \approx 260$ K[29]. A representative linecut of dI/dV conductance at 77 K is shown in Fig. 1f revealing a fully formed CDW gap with gap magnitude $\Delta_0 \approx 200$ meV. We note here that while the material shows a gap even above $T_c$[30–32], the resistance increases dramatically below ~65 K, so STM and point contact were performed at or above 65 K.

Figure 2a shows a schematic of our experimental setup for ultrafast tunneling and point contact measurements (instrumentation and measurement details can be found in Methods and Supplementary Fig. 2, 3). Two identical ultrafast laser pulse trains with delay time $t_d$ (pump #1, #2 – wavelength: 1025 nm, pulse width: 240 fs, repetition rate: 500 kHz) illuminate the junction. The generated ultrafast currents $i_{1,2}(t)$ are time-averaged by a pre-amplifier with 1 kHz bandwidth, giving a current $I(t_d, i_1, i_2)$ at each set delay time, $t_d$. This configuration can resolve ultrafast dynamics from a collective mode in the following manner.

Figure 2b illustrates $i_{1,2}(t)$ at various $t_d$ in a case where the pump pulses excite a collective mode that leads to oscillatory current; $i(t) \sim E \cos \omega t$ ($E$ is laser electric field). When $t_d$ is much larger than relaxation timescale $\tau_r$ of the dynamics ($t_d \gg \tau_r$), each pump #1,#2 provides identical, independent contribution to $I(t_d = \infty) \sim \int i_1 dt + \int i_2 dt$. When $t_d < \tau_r$, one can imagine a small correlation contribution ($i_1 i_2$ to the leading order) so that $I(t_d) \sim \int i_1 dt + \int i_2 dt + \int i_1 i_2 \, dt$. With a lock-in technique (Method Sec. 4), one can extract time-averaged pair-pulse correlation: $\Delta I(t_d) = I(t_d) - I(\infty) = \int i_1 i_2 dt$. As illustrated and derived in Supplementary discussion Sec. 1, the correlation current $i_1 i_2 \sim E^2[\cos \omega t_d + \cos \omega(2t - t_d)]$ consists of an offset determined by $t_d$ (the first term) and an oscillatory part with respect to real time $t$ (the second term). Integration over time constant $T_{int}$ leaves only the offset part. The resulting "transient current" $\Delta I(t_d) \sim E^2 T_{int} \cos \omega t_d$ oscillates in the delay time $t_d$ at the same frequency $\omega$ as the ultrafast current $i_{1,2}(t)$ from the collective mode. In this manner, one can study collective excitation of a system from the transient current $\Delta I(t_d)$. Figure 2d shows the transient current normalized by the setpoint ($\Delta I/I$) obtained from UPC. The data indeed exhibit oscillations as a function of $t_d$ and a Fourier transform of the time-domain data reveals two distinct peaks near 0.11 THz and 0.22 THz repectively.

While ultrafast tunneling also shows clear oscillating transient current, under illumination, the loosely bonded surface iodine atoms can sometimes move which prevents consistent spatially dependent ultrafast tunneling studies in this material (see Supplementary Fig. 4). Materials where the surface atoms are all chemically bonded do not show this migration and consistent ultrafast tunneling under illumination is possible[33]. To overcome this issue, we extend the tip towards the sample from tunneling position, forming a highly controlled point contact with the surface whose lateral diameter is roughly 30 nm (Supplementary Fig. 5). This point contact is impervious to the migration of surface atoms and allows reproducible measurement of ultrafast current oscillation. Additionally, the point contact is more robust for comparing data across different temperatures as it is less sensitive to temperature-sweep induced tip status changes. Henceforth, we focus on the UPC data.

Before proceeding further, we summarize recent findings on $(TaSe_4)_2I$. As previously mentioned, THz emission spectroscopy[28] shows a mode at 0.22 THz, with intensity sharply increasing around 77 K (~$0.3T_c$). Theoretically, massless acoustic phasons result from the coupling of electrons to acoustic phonons at $\mathbf{q}_{CDW}$. However, in insulating materials, when the phason is associated with a longitudinal acoustic (LA) phonon, poor screening of long-range Coulomb interactions, due to the depletion of thermally excited quasiparticles, leads to a predicted band renormalization of the phason to a massive optical band[5,6]. This is expected to happen not at $T_c$ but at a lower temperature where the renormalized phason velocity exceeds velocity of quasiparticles which for this material is approximately $0.3T_c$[6,28]. Consequently, the 0.22 THz mode is thought to originate from a massive phason. This scenario is constrained to LA phasons; transverse acoustic (TA) phasons remain massless even when the Coulomb interactions are not screened. Additionally, time-resolved X-ray measurements[34] of the CDW sideband peaks at $\mathbf{q} = \mathbf{q}_{Bragg} \pm \mathbf{q}_{CDW}$, show oscillations at 0.11 THz. These were interpreted as a low-energy amplitudon associated with an acoustic mode that plays a role in the formation of the CDW[34–36]. We emphasize that the X-ray studies were conducted at 150K, where the 0.22 THz emission observed in THz spectroscopy is no longer detectable. This supports the hypothesis that the modes observed by X-ray and THz emission have distinct origins.

To explore the origins of the UPC modes, we begin by examining their temperature dependence. Figures 3a and 3b show the time and frequency domain profiles of $\Delta I/I$ at selected temperatures. A visual analysis reveals that the intensities (peak heights) of both modes increase as the temperature drops. The mode intensities with temperature are plotted in Fig. 3c, along with the intensity of the 0.22 THz mode from the THz emission study. We find that both UPC modes show a sharp increase in intensity below 77 K (~0.3 $T_c$), mirroring the temperature trend of the 0.22 THz emission mode. This raises an intriguing question: given that there is only one longitudinal acoustic (LA) phonon at $\mathbf{q}_{CDW}$ in this system[35], which should lead to a single massive phason, what do these two modes represent? Additionally, while modes such as the amplitudon or coherent phonons may indeed become sharper and more intense with decreasing temperature, the significant and rapid intensity increases near 0.3 $T_c$ we observe here are not expected for these other modes[37–43].

To gather further insights, we turn to optical pump-probe reflectance (OPPR) measurements, which offer a complementary perspective and have been utilized to detect coherent phonons and amplitudons in various charge density wave (CDW) systems[37–43]. It's important to note that our OPPR measurements were conducted using the same experimental setup and carried out concurrently with the UPC measurements. The experimental configuration is depicted in Fig. 4a, Supplementary Fig. 6, and detailed in the Methods. Figures 4b and 4c display the time and frequency domain optical reflectance at various temperatures. From this data, we identify two well-defined modes at 17 GHz and 0.11 THz, and a weaker mode at 0.22 THz. The 17GHz mode shows different probe polarization ($\mathbf{P}_{probe}$) and temperature dependences compared to the OPPR 0.11 and 0.22 THz modes (Supplementary Fig. 7), suggesting that it has a different character. We therefore stay focused on the 0.11 and 0.22 THz OPPR modes.

Our first objective is to repeat the temperature dependence measurements for the 0.11 and 0.22 THz OPPR modes, as shown in Fig. 4d. Surprisingly, the 0.11 and 0.22 THz OPPR modes exhibit a decrease in intensity below 77 K (~0.3 $T_c$), in contrast with the behavior of the UPC and THz emission modes. This suggests that the modes detected by OPPR and UPC represent different types of coherent excitations, despite their similar frequencies. Based on the divergent temperature dependences of the UPC and OPPR modes as well as the observation of the 0.11 THz amplitudon in prior X-ray studies, we speculate that the 0.11 THz OPPR mode represents the amplitudon, with the much weaker 0.22 THz mode being its second order counterpart. This observation is in agreement with x-ray measurement data which show that the 0.11 THz amplitude mode has a finite amplitude at 150 K, similar to the OPPR mode. One might conjecture here that the divergent temperature dependences of the UPC and OPPR modes could be due to differences in their temperature-dependent detection sensitivity to the same modes. However, such a scenario is unlikely because another optical spectroscopy namely the optical-pump THz-emission study[28], exhibits a completely different temperature dependence with OPPR for 0.22 THz mode (Fig. 4d)). Note that even though the noncentrosymmetric structure of $(TaSe_4)I_2$ enables coupling to phasons, their absence in the OPPR response can be attributed to a small Raman tensor element, as elaborated in Supplementary Discussion Sec. 2. Also note that the comparison of our OPPR results to previous literature[36] based on differences in polarization direction is discussed in Supplementary discussion Sec. 3.

With the OPPR data in hand, we now explore the potential origins of the modes observed in UPC measurements. A straightforward explanation for the 0.11 THz mode in UPC is that it represents the massive phason, with the 0.22 THz mode being its second-order equivalent. However, this explanation is unlikely for two key reasons. First, second-order modes typically have much weaker intensities than first-order modes, yet the intensities of the two UPC modes are almost identical (see Fig. 2c and Fig. 3c, f). Second, no third or fourth-order modes were found at 0.33 and 0.44 THz as shown in Fig. 2d, despite the strong intensity of 0.22 THz mode. Third, as previously noted, X-ray studies[34] have identified an amplitudon at 0.11 THz, making it unlikely that the massive phason has the same energy[6]. Fourth, the 0.22 THz UPC mode shows linear fluence dependence near the value where the measurements are taken (Supplementary Fig. 8), unlike quadratic dependence expected for 2$^{nd}$ harmonic mode.

A more plausible explanation for the 0.11 THz mode in the UPC measurements is that it represents a parametrically amplified excitation of a massless acoustic phason at 0.11 THz, resulting from the decay of the massive optical phason at 0.22 THz. Parametric amplification, as illustrated in Fig. 5a, is a dynamical phenomenon where a massive particle decays into two massless particles moving in opposite directions. In this process, an optical mode at $(k, \omega)$ = $(0, 2\omega_p)$ can parametrically excite acoustic modes at $(-q_p, \omega_p)$ and $(+q_p, \omega_p)$ provided there is anharmonic coupling between them. This mechanism has been theoretically proposed[44] and experimentally observed as the decay of an amplitudon (optical) into massless phasons (acoustic) in blue bronze[45]. We observe that based on this data, the previous THz spectroscopy experiment[28] should have also detected a 0.11 THz emission mode. The absence of this mode in the THz specroscopy can be attributed to the spectrometer's limited sensitivity at these energy scales, as mentioned in the previous THz emission studies[28,46]. In identifying the 0.11THz mode as a parametrically amplified phason, one must ask why we don't measure the amplitudon with UPC. In fact, unlike THz-STM[47], the signature of the amplitudon cannot be captured in UPC. This is because (as illustrated in Supplementary Fig. 9), the current oscillations from an amplitudon have opposite signs within one CDW wavelength, and the ≈30 nm-size point contact averages currents from more than 3 CDW wavelengths. In contrast, for the case of coherent massive phasons, the charge density distribution oscillates as one entity providing detectable oscillations in the point contact current.

This scenario can account for the similar mode intensities between the 0.11 and 0.22 THz UPC modes, as well as their consistent frequency ratio locking over temperature and bias (Supplementary Fig. 10). It necessitates a massless phason band with transverse acoustic (TA) character, which has been observed in previous microwave studies[3,48]. Moreover, since the intensity of the 0.11 THz mode directly depends on the 0.22 THz mode, it naturally explains their similar temperature dependence and $\mathbf{P}_{pump}$ dependence (Supplementary Fig. 11), despite the 0.11 THz mode exhibiting TA character. Finally, this scenario aligns with the temperature dependence of the OPPR modes, which we will discuss next.

The unusual temperature dependence observed in our OPPR data manifests below 77 K, where the intensities of both the 0.11 THz and 0.22 THz modes decrease as temperature drops. This behavior contrasts with findings in other CDW systems[37–43], where the scattering rate typically decreases and the CDW amplitude increases upon cooling. Moreover, the temperature range in which the OPPR modes begin to diminish coincides with the range where phasons from UPC measurements at similar frequencies start to exhibit significant intensities. Based on all our data, we propose the following scenario: when an optical pump excites the sample, hot electrons alter the free-energy landscape, shifting it from the dotted curve to the solid curve in Fig. 5b. This process activates the displacive excitation of amplitudons[49] at 0.11 THz (seen in our OPPR studies; red arrow), As the temperature decreases below 0.3 $T_c$ (Fig.

5c), quasiparticle depletion leads to an unscreened long-range Coulomb interaction, which introduces curvature in the free-energy landscape along the phase axis. Under optical excitation, the position of the free-energy minimum shifts in phase coordinate due to a coupled phonon[28], thereby activating a 0.22 THz massive phason (observed as current oscillations in the UPC; blue arrow). The massive phason then parametrically excites a 0.11 THz massless transverse phason which is also seen as current oscillations, and eventually both modes radiate away (magenta curve). As both phason mode intensities sharply increase at low temperature, less energy is available for exciting the amplitudon at 0.11 THz, decreasing its intensity as seen in the OPPR measurements. This results in the suppression of the amplitudon intensity in the OPPR measurements. This type of strong coupling between two modes at similar frequencies has been previously observed in various materials utilizing techniques such as Raman and neutron scattering[50–53]. Overall, this scenario illustrates how the massive phason interacts with other collective excitations within a CDW system.

In summary, we have developed a multimodal ultrafast STM-based setup that allows direct measurements of oscillating currents at femtosecond timescales. By utilizing point contact, we have detected oscillating currents associated with a massive phason, along with a parametrically excited massless phason at half that frequency, resulting in a second oscillating mode. This parametric amplification resembles the decay of a neutral pion into two photons, and its competition with the amplitudon observed in our optical reflectivity measurements underscores the complex dynamics within a charge density wave system. Our research provides new insights for exploring not only the dynamics of the CDW state but also the local dynamics of interesting quantum materials at ultrafast timescales, including light-induced superconductivity[54] and the light-induced anomalous Hall effect[55].

**Acknowledgements**

This material is based on work supported by the US Department of Energy Office of Science National Quantum Information Science Research Centers as part of the Q-NEXT center, which supported the laser-STM work of S.B. and A. R. and provided partial support for laser-STM development. V.M. acknowledges support from the Gordon and Betty Moore Foundation's EPiQS initiative through grant number GBMF9465 for the laser-STM instrument development. Z.W. is supported by the US Department of Energy, Basic Energy Sciences (grant number DE-FG02-99ER45747). Crystal growth was supported by the Center for Quantum Sensing and Quantum Materials, an Energy Frontier Research Center funded by the U. S. Department of Energy (DOE), Office of Science (SC), Basic Energy Sciences (BES) under Award DE-SC0021238. B.B. acknowledges support from the National Science Foundation under Grant No. DMR-1945058. The authors are immensely grateful to D. Cahill, E. Berg, S.L. Cooper, A. Cavalleri, E. Baldini, J. Demsar, and N. Wang for useful discussions.

**Author Contributions:**

S. B., and V. M. conceived the project. S. B. and A.R. constructed the UPC and OPPR setup and conducted the UPC and OPPR measurements. S.K. and F.M. conducted THz-emission measurement. K. Q., C.Z., and D.P.S. provided $(TaSe_4)_2I$ samples used in this study. Z.W., F. M. and B.B. contributed to the theoretical interpretation. S.B., A.R., and V. M. performed data analysis and wrote the manuscript with input from all the authors.

**Competing interests:** The authors declare that they have no competing interests.

**Data Availability:** Data for the main figures and Supplementary figures will be available at the Illinois Databank.

**Figure 1**

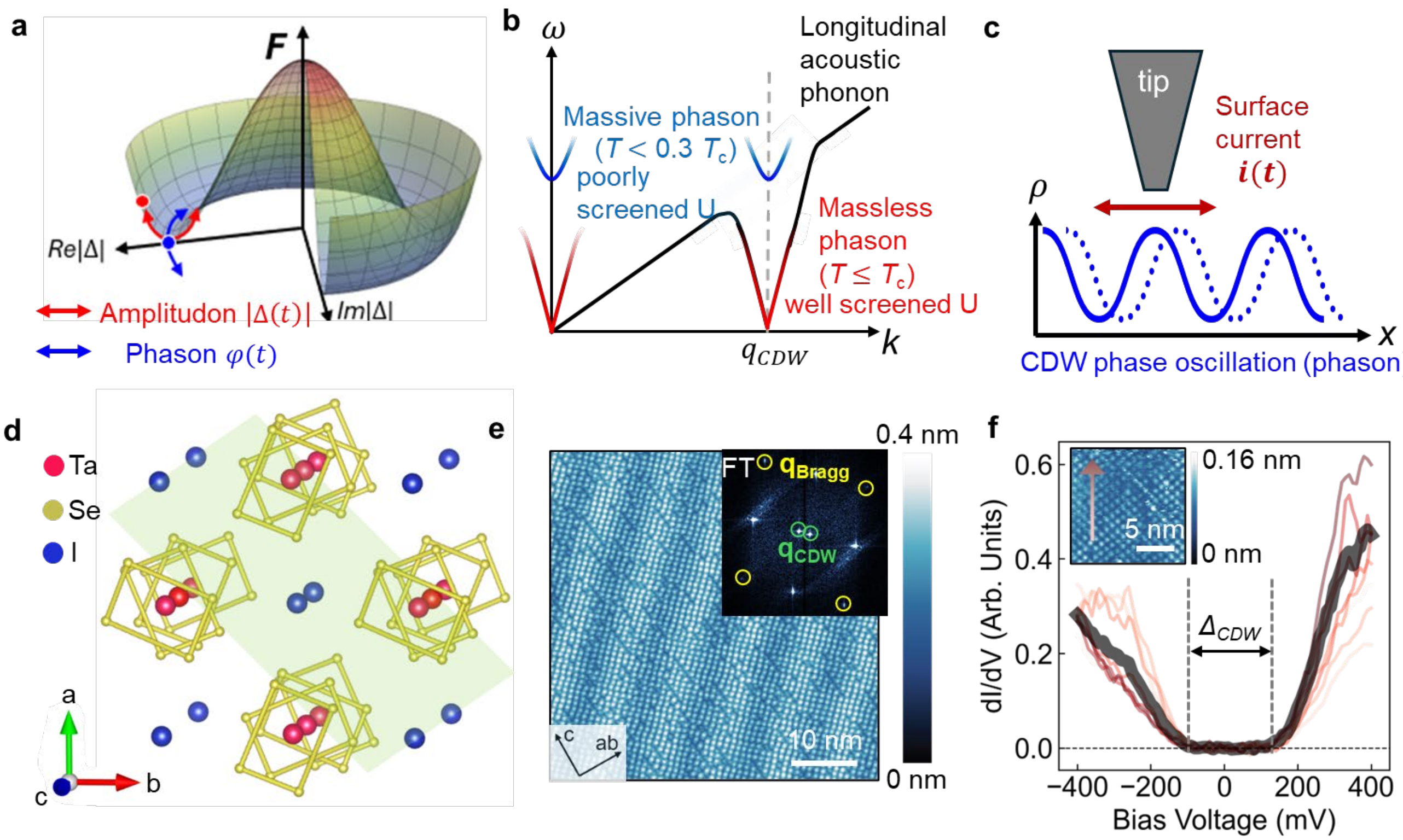

**Fig. 1 | Collective modes of charge density wave state (CDW), crystal structure of $(TaSe_4)_2I$, and STM characterization.**

**a,** Free energy diagram of CDW state showing two types of collective excitations. The red/blue curves denote the amplitude mode (amplitudon)/phase mode (phason) where the magnitude/phase of the charge density oscillates. **b,** Schematic dispersion relation of phonon and phason bands for two different regimes of the Coulomb interaction (U). Below the CDW transition $T_c$, a longitudinal acoustic (LA) phonon that couples to the CDW at $\mathbf{q}_{CDW}$, softens into a massless phason band (red) which gets band folded to zero momentum. At high temperatures near $T_c$, thermally excited quasiparticles screen long-range U. Upon cooling ($T < 0.3\ T_c$), the quasiparticles freeze and consequently, long-range U is poorly screened. This can result in a spectral shift of phason band up to optical frequencies resulting in a massive phason. **c,** Schematic of charge oscillations resulting from exciting the phason. The oscillating surface current can be measured by ultrafast tunneling/point contact measurements. **d,** Crystal structure of $(TaSe_4)_2I$, showing a quasi-1D structure with the $TaSe_4$ chain direction along the *c*-axis. The crystal cleaves at the (110) plane resulting in an iodine termination (light green plane). **e,** Topographic image of the cleaved surface at bias voltage $V_{bias}$ = -1.00 V and setpoint $I$ = 2.50 pA at 77 K showing the iodine lattice. The inset shows Fourier transform of (e) which displays CDW wavevector $\mathbf{q}_{CDW}$ (green circles) and Bragg peaks $\mathbf{q}_{Bragg}$ (yellow circles). **f,** dI/dV spectra along a line in the topography image in the inset at T = 77 K ($V_{bias}$ = -1.00 V, $I$ = 50.0 pA). The black curve is an average of the 8 individual spectra in the linecut.

**Figure 2**

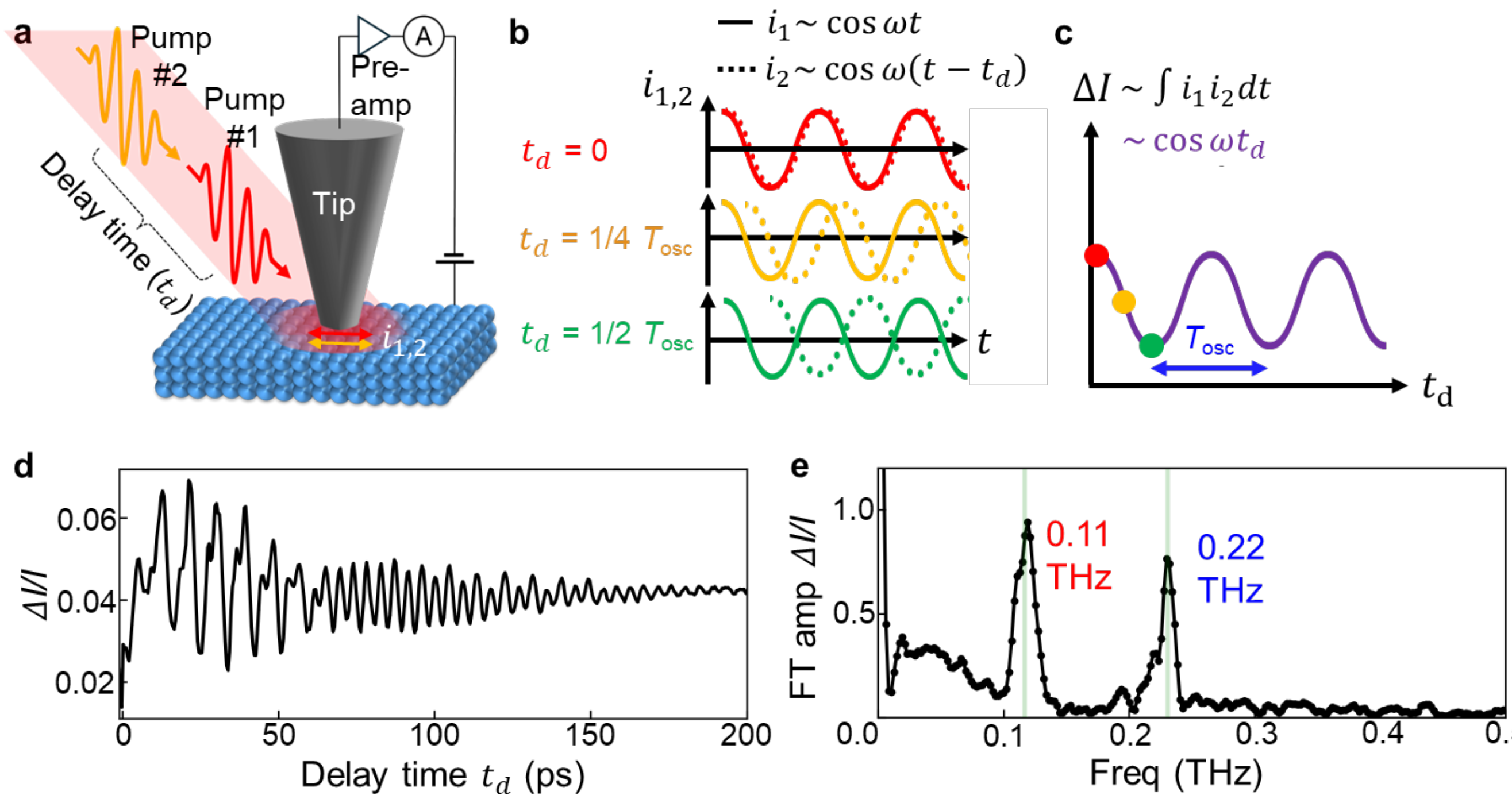

**Fig. 2 | Light-induced oscillating currents observed by ultrafast point contact (UPC) current measurements**.

**a,** Schematic of UPC and ultrafast tunneling setups. The STM tip (gray) is in contact or in tunneling with respect to the sample (blue). The red and orange wavy lines denote two identical optical pump pulses (pump #1, #2), which illuminate the junction with a delay time ($t_d$) between them. $i_1(t)$ *and* $i_2(t)$ are generated ultrafast currents due to pump #1 and pump #2, respectively. **b**, $i_1$ (solid curves)and $i_2$ (dotted curves) at various delay times $t_d$ for the case where the pump excites a collective mode that leads to oscillatory current; $i(t) \sim E \cos \omega t$ ($E$ is laser electric field). $T_{osc}$ = $2\pi/\omega$ is the oscillation period of $i(t)$. **c**, The correlation current $i_1 i_2$, which contains information about system's dynamics, when $t_d$ is shorter than the relaxation timescale of the collective mode. $i_1 i_2 \sim E^2[\cos \omega t_d + \cos \omega(2t - t_d)]$ consists of an offset (first term) determined solely by $t_d$, and oscillatory part (second term) with respect to real time $t$. We measure the time averaged term $\int i_1 i_2 \, dt$ which zeros the second term, so that $\int i_1 i_2 \, dt \sim E^2 T_{int} \cos \omega t_d$, where $T_{int}$ is an integration time. $\int i_1 i_2 \, dt$ at various $t_d$ can be extracted by measuring the "transient current" $\Delta I(t_d) = I(t_d) - I(\infty)$ with lock-in technique (Method Sec. 4). $\Delta I$ oscillates with $t_d$ at the same frequency $\omega$ as the collective mode (for more detailed derivation, see Supplementary discussion Sec. 1). **d**, **e,** Experimentally measured time domain ($t_d$) and frequency domain transient UPC current $\Delta I$ normalized by setpoint *I* at 77 K. $V_{bias}$ = - 10 mV, *I* = 26.9 pA and the tip extension ($Z_{ext}$) = 25 nm from the tunneling regime ($V_{bias}$ = -1V, *I* = 20 pA). Each pump has fluence (FL) = 0.145 mJ/cm$^2$ and pump polarization $\mathbf{P}_{pump}$ ∥ c-axis. Modes at ≈ 0.11 and 0.22 THz (vertical green lines in (e)) are visible in both time and frequency domains.

**Figure 3**

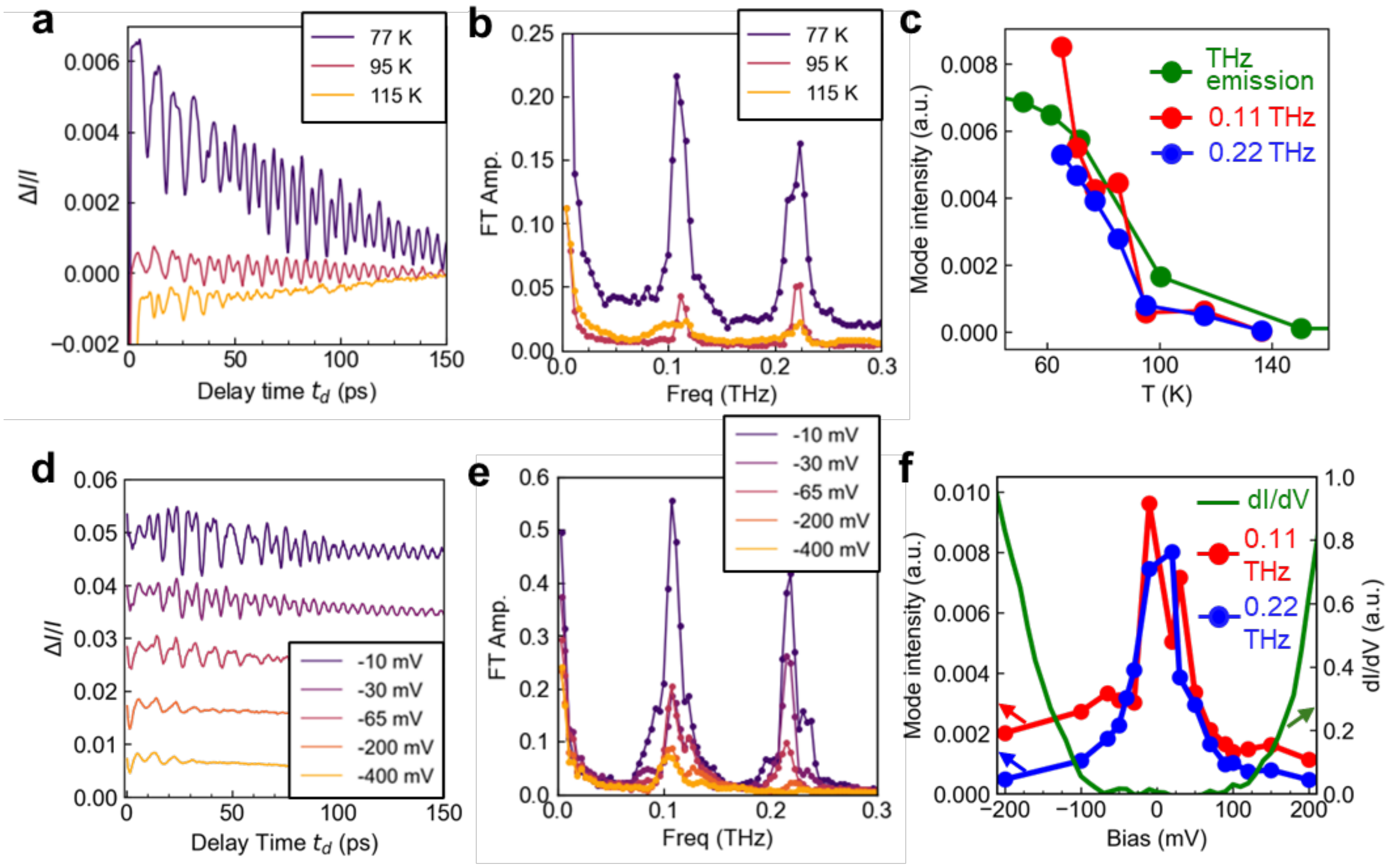

**Fig. 3 | Temperature and bias dependence of THz modes in UPC measurement**

**a**, Normalized transient current $\Delta I/I$ at representative temperatures obtained from UPC measurements in the time domain ($V_{bias}$ = - 10 mV, $Z_{ext}$ = 25 nm). See Methods Sec. 3 for the polarization, fluence, and setpoint setting for (a-c). **b**, Angle averaged FT of the $\Delta I/I$ showing the oscillating current modes at 0.11 and 0.22 THz. **c**, The intensities of 0.11 THz (red) and 0.22 THz (blue) UPC modes, overlayed with the intensities of the 0.22 THz mode in the recent optically excited THz emission study[28] (rescaled). All three modes display similar temperature dependence with a sharp rise below 77 K ($\approx$ 0.3 $T_c$). **d**, UPC $\Delta I/I$ at representative biases at time domain ($T$ = 77 K, $Z_{ext}$ = 80 nm, FL = 0.145 mJ/cm$^2$, $\mathbf{P}_{pump}$ ∥ c-axis, see Methods Sec. 3 for setpoints). **e**, FT of (d). **f**, Bias dependence of UPC mode intensities (red, blue) overlayed on the dI/dV conductance spectrum (green). The mode intensities start to rise sharply within the CDW gap $\approx \pm$100mV seen in the dI/dV spectrum.

**Figure 4**

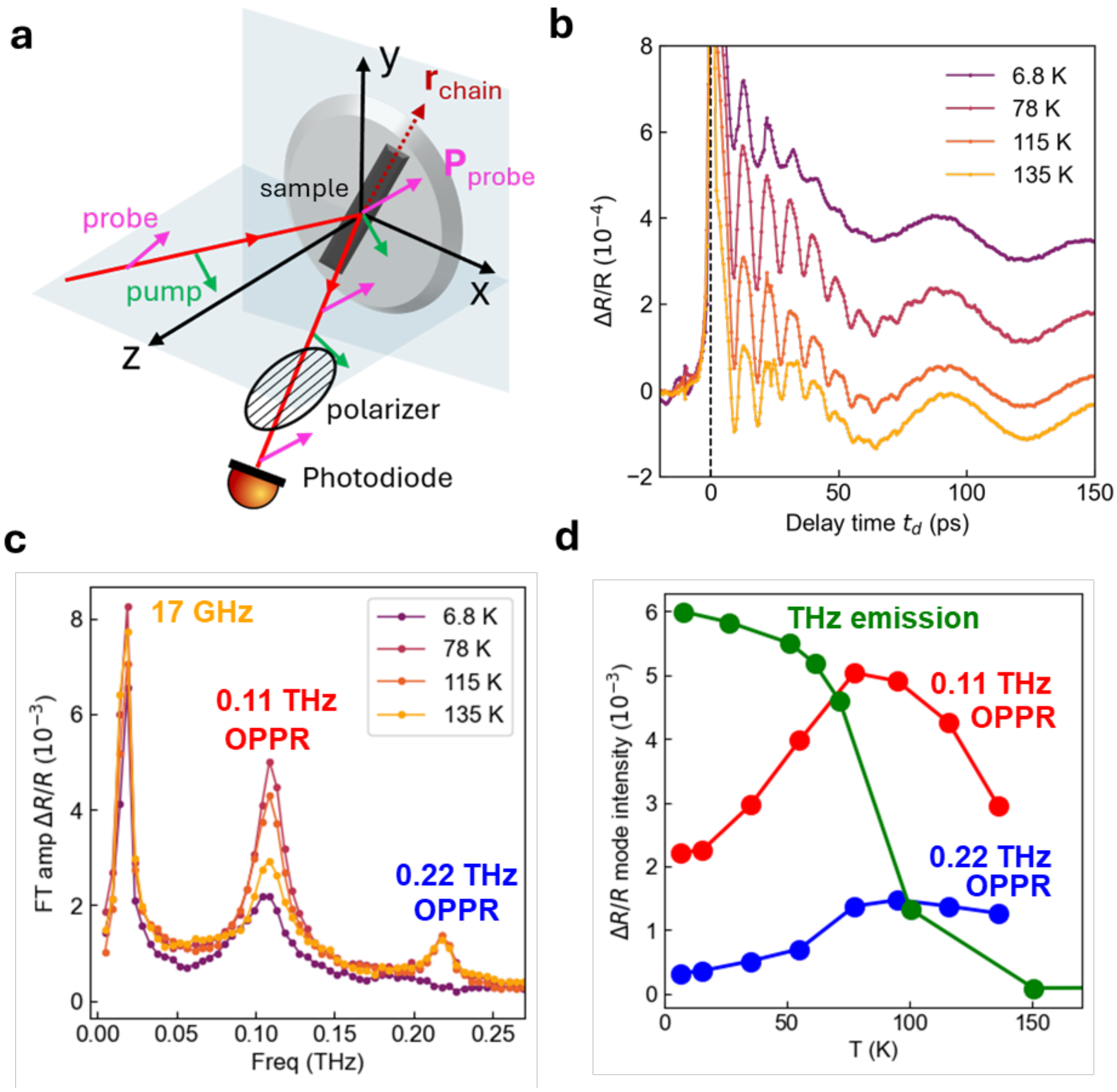

**Fig. 4 | Temperature dependence of the optical pump-probe reflectance (OPPR).**
**a**, A schematic of OPPR setup. $\mathbf{P}_{pump}$ and $\mathbf{P}_{probe}$ are kept orthogonal to each other, and the pump is filtered by polarizer in front of the photodiode. **b**, Normalized OPPR (*ΔR/R*) at various temperatures angular averaged over $\mathbf{P}_{pump}$ from 0 to 180 degrees. The pump FL was fixed to 0.174 mJ/cm$^2$. **c**, The FT of OPPR averaged over $\mathbf{P}_{pump}$. 3 modes are identified at 17 GHz, 0.11 THz, and 0.22 THz. **d**, Temperature dependence of the mode intensities of 0.11 (red) and 0.22 THz (blue) OPPR modes, along with 0.22 THz mode intensity from the recent THz emission study[28] (green). Temperature dependence of 17 GHz mode can be found in Supplementary Fig. 8b.

## Figure 5

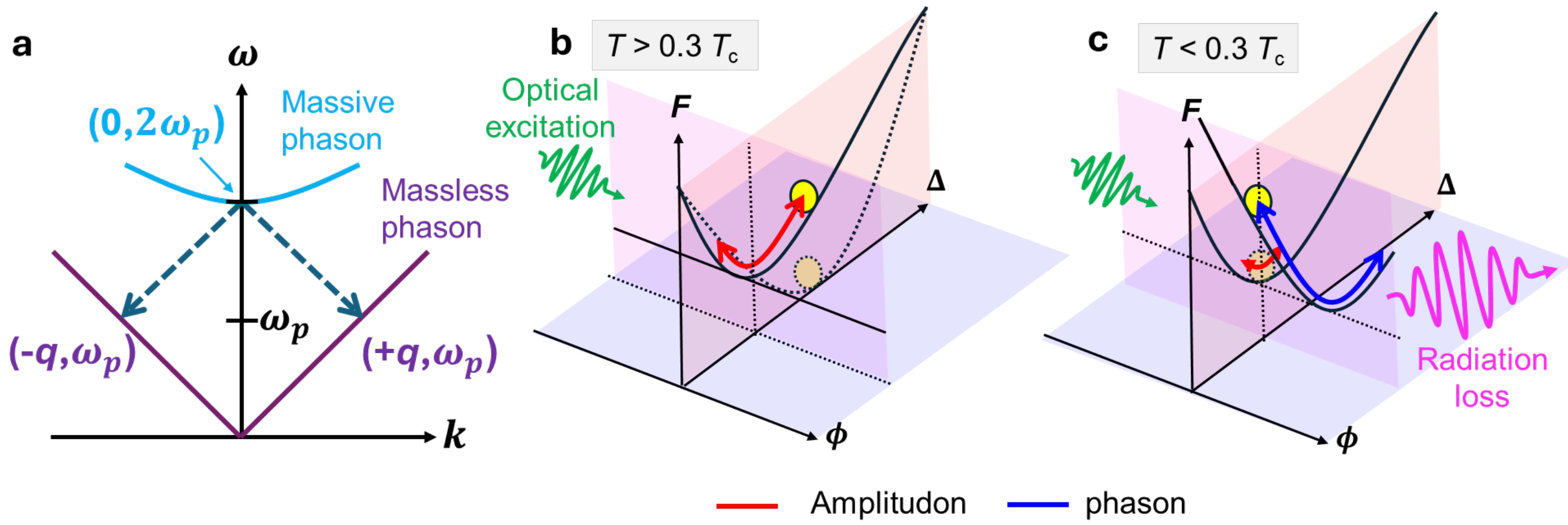

**Fig. 5 | Illustration of the massive-to-massless parametric amplification of phason and the interaction between amplitudon and phason.**

**a**, A schematic diagram for parametric amplification from the massive phason (sky blue) at $(k,\omega)$ = $(0,2\omega_p)$ to the massless phason (purple) at $(-q,\omega_p)$ and $(+q,\omega_p)$ through anharmonic coupling. **b**, Illustration of interaction between the amplitudon and phason. Here, $\Delta$ and $\phi$ are the amplitude and phase of the CDW order parameter. Upon optical excitation (green curve), the hot carriers change the free energy landscape from the dotted line to the solid line. As the system (yellow ball) is no longer located at the free-energy minimum, the amplitude of CDW oscillates (amplitudon; red arrow, measured by OPPR). **c**, At low temperatures ($T$ < 0.3 $T_c$), quasiparticle depletion leads to an unscreened long-range Coulomb interaction, which introduces curvature in the free-energy landscape along the phase axis. Upon optical excitation, the position of the free-energy minimum shifts in phase coordinate due to coupled phonon[28], thereby activating a 0.22 THz massive phason (blue arrow), which then parametrically excites a 0.11 THz massless transverse phason (not displayed here). The current oscillations of the phasons result in electromagnetic radiation (magenta curve), dumping the energy from the optical excitation. As the phason intensity sharply increases at low temperature, more energy from optical excitation goes to exciting the phason and the corresponding radiation, leaving less energy to excite the amplitudon, thereby suppressing the amplitudon intensity.

# Methods and Supplementary information for

## Dynamic competition between phason and amplitudon observed by ultrafast multimodal scanning tunneling microscopy

**Table of Contents**

## Methods

### 1. Single crystal growth of $(TaSe_4)_2I$

Single crystals of $(TaSe_4)_2I$ used in this study were grown by vapor transport. The detailed recipe, transport and X-ray characterization can be found in Reference[1].

### 2. Scanning tunneling microscopy/spectroscopy

The $(TaSe_4)_2I$ samples used in the experiments were cleaved in situ at room temperature and immediately transferred to an STM chamber. All STM experiments were performed in an ultrahigh vacuum (better than $1 \times 10^{-10}$ Torr). All the parameters (setpoint voltage and current) of the STM topographic images and dI/dV conductance spectra are listed in the captions of the figures. Unless otherwise noted, the differential conductance ($dI/dV$) spectra were acquired by a standard lock-in amplifier at a modulation frequency of 913.1 Hz. The STM tip made from tungsten was fabricated by electrochemical etching.

### 3. Ultrafast point-contact (UPC) measurement

The point-contact is made by extending the Z position of the tip towards the sample for 25 nm (otherwise mentioned) from the tunneling regime ($V_{bias}$ = -1V, $I$ = 20 pA). The feedback loop is on to keep the constant DC current during the delay time sweep. As shown in the optics layout of the UPC setup (Supplementary Fig. 2), the ultrafast optical pulse train of 1025 nm wavelength (1.21 eV), 250 fs pulse width, and 1 MHz repetition rate is generated by Yb:KGW regenerative amplifier laser system (PHAROS PH2-10W, Light Conversion). The generated pulse train is split into two by 50:50 beam splitter (pump #1 and pump #2). For a lock-in measurement of pair-pulse correlation to extract the transient point-contact current $\Delta I$ (See the next section of Methods), each pulse train passes through an individual acousto-optic modulator (AOM) which picks every even or odd pulses in the train, so that the repetition rate of the output becomes 500 kHz. The pulse picking timing is controlled by the waveforms sent from an arbitrary function generator (AFG) (ARB Rider 2182, Active Technologies). A lock-in amplifier sends TTL trigger whose state switches the AFG and AOM between even- to odd-pulse-picking mode, giving a digital modulation on the delay time $t_d$. Then pump #2 goes through mechanical delay stage to gives a set delay time ($t_d$). The two pulse trains are merged by beam splitter and become colinear. Before getting into the STM chamber, the polarizations of the pump #1, #2 are rotated together by a half-wave plate to align them to the specific crystallographic direction of interest. Then, both pumps pass through $CaF_2$ view port and are focused on the point-contact by a ZnSe aspheric lens (Avantier Inc.) on a piezo stage of USM1200LL STM (Unisoku). The incident angle of the beam to the surface normal direction of the sample is 55 degrees. The beam is focused to an elliptic spot (due to the incidence angle) with a size estimated to be 10 $\mu$m x 17 $\mu$m (1/$e^2$ diameter). To ensure the alignment of the beam spot to the point-contact, the reflected beams go to a CMOS camera. The measured current from the contact during illumination $i(t)$ is amplified by a pre-amplifier (Femto DLPCA-200) and goes to STM controller to run the feedback loop. The transient component of the current $\Delta I(t_d)$, which is modulated at lock-in TTL frequency, is demodulated at the lock-in amplifier.

For data shown in Fig. 3a-c of the main text, $\mathbf{P}_{pump}$ was swept from 0 to 180 degrees with 15-degree intervals and the $\Delta I/I(t_d)$ was averaged over $\mathbf{P}_{pump}$. The FL when $\mathbf{P}_{pump}$ ∥ c-axis is 0.145 mJ/cm$^2$ (same as Fig. 2c) and the FL was adjusted for other $\mathbf{P}_{pump}$ so that laser in-plane electric field was kept the same during the sweep. The setpoint at each $\mathbf{P}_{pump}$, temperature, and bias for data shown in Fig. 3a-c and 3d-f can be found at the attached data summary spreadsheet for referee and will be uploaded to Illinois Databank due to its large volume of table.

## 4. Lock-in measurement of the pair-pulse correlation for extracting transient current

The concept of lock-in measurement of the pair-pulse correlation for extracting transient current is introduced in Ref[2–4], but it is explained here again for the reader's convenience. As shown in Fig. 2b of the main text, correlation of the ultrafast currents $i_1(t)$, $i_2(t)$ generated by two pump pulses (pair-pulse correlation) could introduce delay time $t_d$ dependent time($t$)-averaged total current $I(t_d)$. Because the transient component ($t_d$ dependent) due to the pair-pulse correlation is typically much smaller compared to the DC component, the transient component needs to be separated from DC component and noises at other frequencies to make a sensitive measurement on it with a detectable signal-to-noise ratio. To separately extract the transient component due to pair-pulse correlation ($\Delta I$) from the $I$, a lock-in amplifier digitally modulates the delay time between two pump pulses in following manner.

As illustrated in Supplementary Fig. 3, the pulse trains of pump #1,#2 at 1 MHz pass through AOM #1, #2 which act as pulse picker #1, #2. The pump #2 passes through the delay stage which imposes a short set delay time $t_d$ < 1 ns. Then, both pumps are merged by a beam splitter and illuminate the sample. In terms of pulse picking, the pulse picker #2 always picks even pulses and blocks odd pulses regardless of lock-in TTL trigger state. On the other hand, the pulse picker #1 picks even pulses when the lock-in TTL volage is high, and picks odd pulses when the TTL voltage is low. As a result, when the TTL voltage is high, the delay time between pump #1,#2 becomes the set delay time $t_d$ set by the delay stage. If $t_d$ is smaller than the relaxation time of the dynamics of the system $\tau_r$ upon optical excitation, the $t_d$ dependent pair-pulse correlation, which tells us about the dynamics of the system, will be included in the time-averaged current $I(t_d)$. Now, when the TTL voltage is low, the delay time between pump #1, #2 becomes $t_d$ + one pulse interval (1 / 1 MHz = 1 $\mu$s). As 1 $\mu$s is typically much longer than the relaxation time $\tau_r$ of coherent dynamics in most systems, the pair-pulse correlation contribution in the $I(t_d)$ is gone.

Once the lock-in starts to modulate TTL voltage between high and low state, the delay time between pump #1, #2 digitally modulates between $t_d$ and $t_d$+∞ (1 $\mu$s), resulting in a modulation of time-averaged current $\Delta I(t_d) = I(t_d) - I(\infty)$. This modulation $\Delta I(t_d)$ only extracts the $t_d$ dependent transient component of the current due to pair-pulse correlation, providing picosecond time-resolution in UPC measurement. Note that the modulation frequency (477 Hz) is lower than the preamplifier bandwidth (1kHz) so that the extracted transient UPC current $\Delta I(t_d)$ can be measured even after the preamplification. Also note that the pairs of pulses come at 500 kHz, repeating them ~ 500 times during each high and low state of the lock-in TTL voltage. This accumulates the small signal from the individual pair-pulse correlation into measureable level.

**5. Optical pump-probe reflectance measurement (OPPR)**

The OPPR measurement shares the same laser module setting (1 MHz repetition rate, 250 fs pulse width, and 1025 nm wavelength), optical configuration inside the STM chamber (the incident angle of the beam), and pulse timing control scheme (the delay time modulation scheme imposed by the lock-in amplifier, AFG, and AOMs) with the UPC measurement (see Methods Sec. 3, 4). The difference in the OPPR measurements is that the two pulses represent pump and probe beams instead of two pumps. As illustrated in Supplementary Fig. 6, polarizations of the pump and the probe beams are individually controlled by the two half-wave plates so the relative angle between the pump and probe polarizations is kept at 90 degrees in all OPPR measurements presented in this work. After the pump and the probe beams are merged by a beam splitter, they follow a colinear path. The probe pulse arrives at the sample with a delay time $t_d$ with respect to the pump pulse. The probe is reflected from the sample and propagates to the photodiode. A linear polarizer in front of the photodiode blocks the pump beam and transmits the probe beam, so that the photodiode measures reflectance *R* of the probe beam only. During the measurement, the STM tip is retracted > 200 $\mu$m from the surface so that it does not affect incoming and reflected beams.

The pump beam introduces hot carrier dynamics and coherent excitations of the sample, which may affect the band structure of the sample, and is captured by the probe reflectance $R(t_d)$. The lock-in measurement with delay time modulation (see Methods Sec. 4) provides the transient component of the probe reflectance $\Delta R(t_d) = R(t_d) - R(\infty)$.

## Supplementary discussions

### 1. Relation between time-averaged transient current $\Delta I(t_d)$ and real-time ultrafast current $i_{1,2}(t)$.

When two identical pump pulses with delay time $t_d$ illuminate the tip-sample contact, an ultrafast current $i(t) = f(i_1, i_2)$ is generated on top of the set-point current from the DC bias where $i_{1,2}$ are contributions from pump #1 and #2. The STM circuit measures time-averaged current $I = \int [i(t) + i_{set}]dt = \int i(t)dt + I_{set}$. When the delay time $t_d$ is larger than the relaxation time of the system $\tau_r$, each pump generates identical amount of ultrafast current so it can be written that $i_1 = i_2$ and $I(t_d = \infty) = I_{set} + \int [i_1]dt + \int [i_2]dt$. When $t_d < \tau_r$, there comes a small correlation component between $i_{1,2}$. In this case, $I(t_d) = I_{set} + \int [i_1]dt + \int [i_2]dt + \int [i_1 i_2]dt$ (to the leading order). If one obtains the transient component of the time-averaged current $\Delta I(t_d) = I(t_d) - I(\infty)$ from the lock-in technique of delay time modulation (Method Sec. 4), it represents pair-pulse correlation component: $\Delta I(t_d) = \int i_1 i_2 dt$ to the leading order.

Now, let's consider how this $\Delta I(t_d)$ looks when ultrafast current arises from oscillating collective modes excited by pumps in a form of $i_1 = E\cos(\omega t)$ and $i_2 = E\cos(\omega t - \phi_d)$. Here, $\phi_d = \omega t_d$ and $E$ is proportional to the peak electric field of the laser. Then, $\Delta I \sim E^2 \int \cos(\omega t)\cos(\omega t - \phi_d)dt = E^2 \int [\cos(\phi_d) + \cos(2\omega t - \phi_d)]dt = E^2[T_{int}\cos(\phi_d) + \int \cos(2\omega t - \phi_d)\,dt] \approx E^2 T_{int}\cos(\phi_d) = E^2 T_{int}\cos(\omega t_d)$, as the term $[\int \cos(2\omega t - \phi_d)\,dt]$ goes nearly to zero upon time-averaging (Supplementary Fig. 12). Here, $T_{int}$ is the integration time in millisecond timescale. As one can see, the "transient current" $\Delta I(t_d) \sim E^2 T_{int} cos(\omega t_d)$ oscillates with respect to $t_d$ with the same frequency $\omega$ to the individual ultrafast current oscillation from the collective modes excited by pumps. In this way, we suggest that it is possible for the $\Delta I(t_d)$ from UPC measurement to capture ultrafast oscillating current from the collective modes. Also, as $E^2 \sim FL$ (fluence), this model predicts $\Delta I$ from the collective modes is linearly proportional to FL, which is observed in low FL regime (see Supplementary Fig. 8).

### 2. Insensitivity of phasons in OPPR

When a system is inversion symmetric (centrosymmetric crystal), OPPR is sensitive to Raman-active modes and insensitive to infrared-active (IR-active) modes. $(TaSe_4)_2I$ breaks inversion symmetry with its chiral crystal structure, and hence this selection rule does not have to hold. In this case, one may wonder in what way the massless and massive phasons in $(TaSe_4)_2I$ are not probed in OPPR (note that both phasons are IR-active). It turns out that for an IR-active mode to be clearly leaked into OPPR, not only symmetry restriction needs to be lifted, but also the mode needs to have large magnitude of Raman tensor elements[5,6]. It is possible for IR-active phasons to show too negligible leakage to be detected in OPPR if their Raman tensor is much smaller than that of the Raman-active modes such as amplitudon, which is possible considering the configuration of the displacements of each mode. The negligible leakage of some of IR-active modes into Raman spectroscopy has been observed for example in $NdGaO_3$, which breaks inversion symmetry below 200 K[7] and non-centrosymmetric $Bi_{0.21}La_{0.91}Sc_{2.88}(BO_3)_4$[8].

**3. Comparison of our OPPR and previous OPPR work on $(TaSe_4)_2I$**

The previous OPPR work on this material[9] displays no suppression of 0.11 THz OPPR mode intensity (amplitudon) at low temperatures in contrast to our result. Furthermore, the previous work did not detect 17 GHz mode, whereas we observe this mode across the entire temperature range of measurement. A possible systematic origin for these different results may be the difference in measurement geometry and pump polarization. In previous study, the pump beam was incident normal to the surface and s-polarized, thus resulting in no out-of-plane field component. In our OPPR work, the pump beam was incident obliquely (55 degrees from surface normal due to the presence of STM tip stage), introducing a substantial out-of-plane field component similar to the p-polarized configuration. Although the precise mechanism by which measurement geometry and out-of-plane field could influence temperature dependence requires future investigation, we note that 0.22 THz emission, believed to be the signature of massive phason[10], vanishes under s-polarized pump, whereas the emission is strong under p-polarized pump (Supplementary Fig. 13).

In addition to this polarization difference, a possible sample-related origin may also contribute to the difference in OPPR results. Unlike the initial THz emission study[10], a subsequent independent THz emission study[11] using samples from different grower reports no 0.22 THz emission, raising a possibility that the presence of the massive phason may depend on detailed sample growth parameters which affect residual charge carrier density (e.g., from vacancies). In such cases, the suppression of amplitudon (0.11 THz OPPR mode) may not occur as the previous OPPR study reported.

Lastly, one may wonder whether the difference in beam wavelength could give rise to the discrepancy in the OPPR results. To examine whether the observed suppression in our measurement originates from wavelength-dependent material properties associated with the 1030 nm pump and probe, we performed OPPR using a 515 nm pump and probe. The 0.11 THz OPPR mode intensity measured with the 515 nm beam shows a similar low-temperature suppression as that observed with the 1030 nm beam, which rules out this possibility.

**Supplementary References**

**Supplementary Fig. 1**

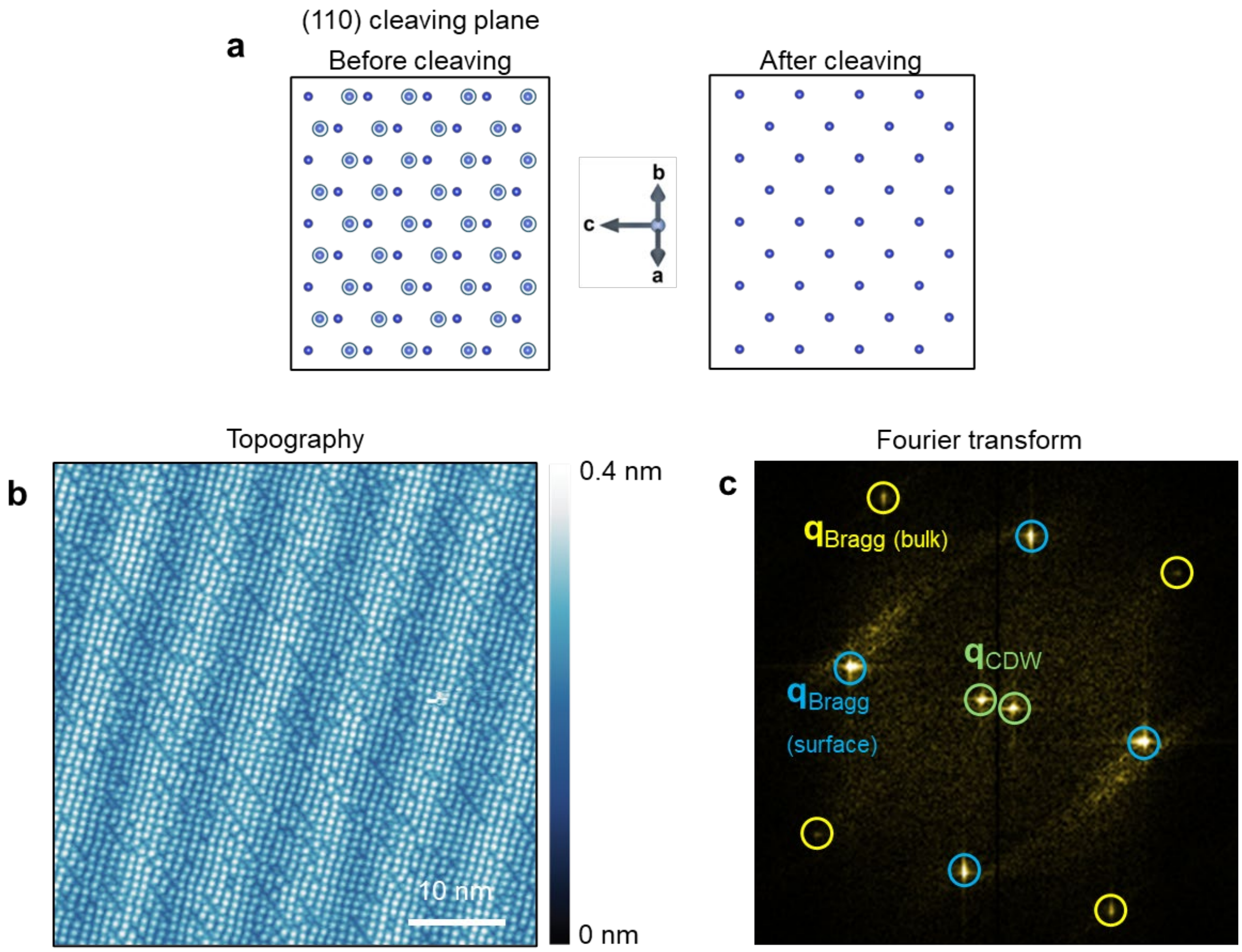

**Supplementary Fig. 1 | Cleavage plane, STM topography, and Fourier transform.**

**a,** Top view of (110) cleaved plane exposing the iodine surface. Approximately half of the iodine atoms are ejected from the surface upon cleaving, leaving behind a square lattice of iodine atoms visualized in STM topographies. **b,** STM topography at $V_{bias}$ = -1.00 V and $I$ = 2.50 pA with atomic resolution of the iodine lattice on the (110) surface along with the incommensurate CDW with the wavelength ≈ 9.1 nm. The topography was taken at a temperature of 77 K. **c,** Fourier transform of the topography in (b) showing bulk Bragg peaks[12], surface Bragg peaks (of iodine lattice), and the CDW peaks.

**Supplementary Fig. 2**

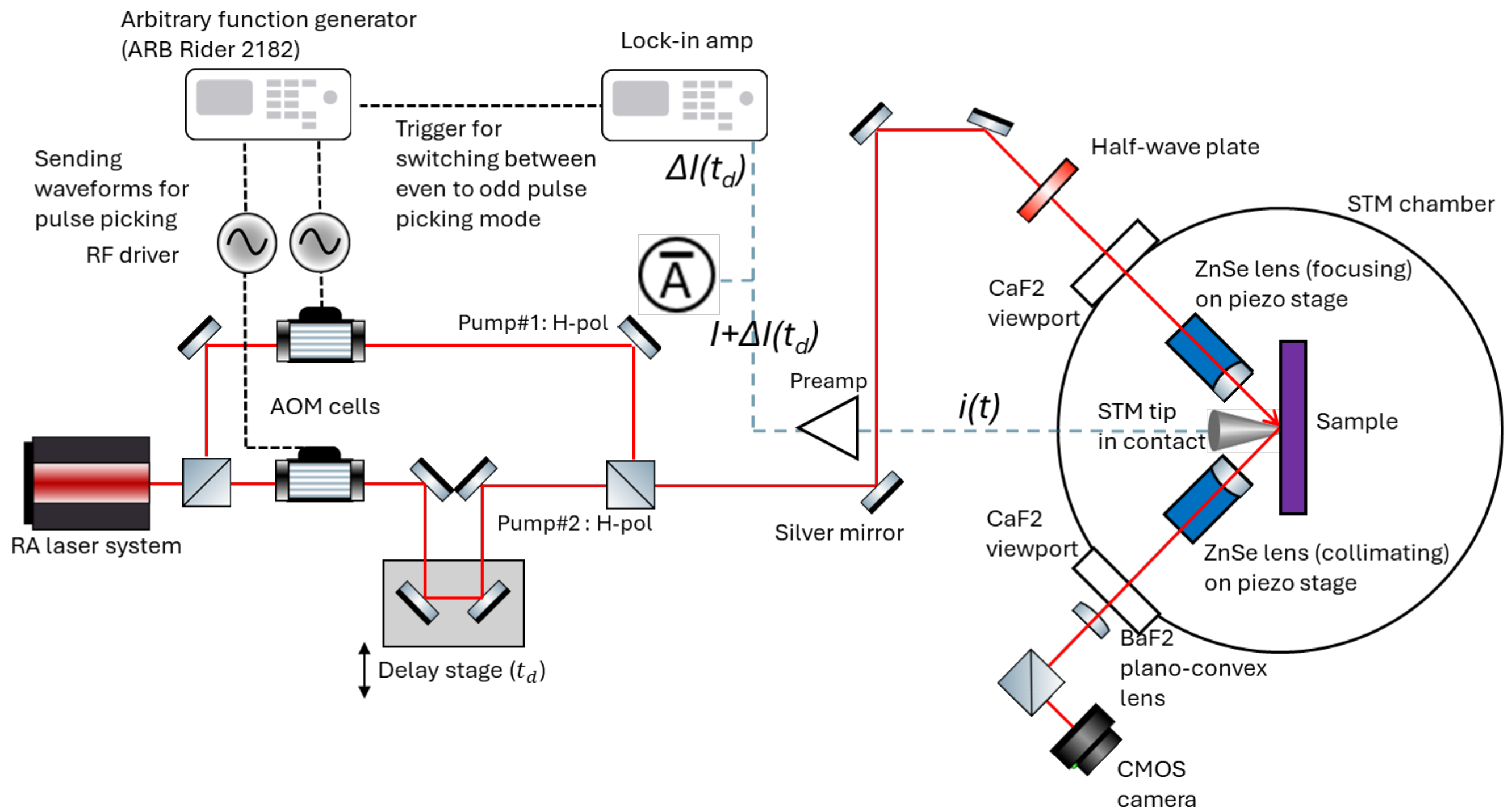

**Supplementary Fig. 2 | Optics layout for ultrafast point-contact current measurement with pair-pulse correlation scheme.**

Detailed explanation of each component can be found in Methods Sec. 3-4.

## Supplementary Fig. 3

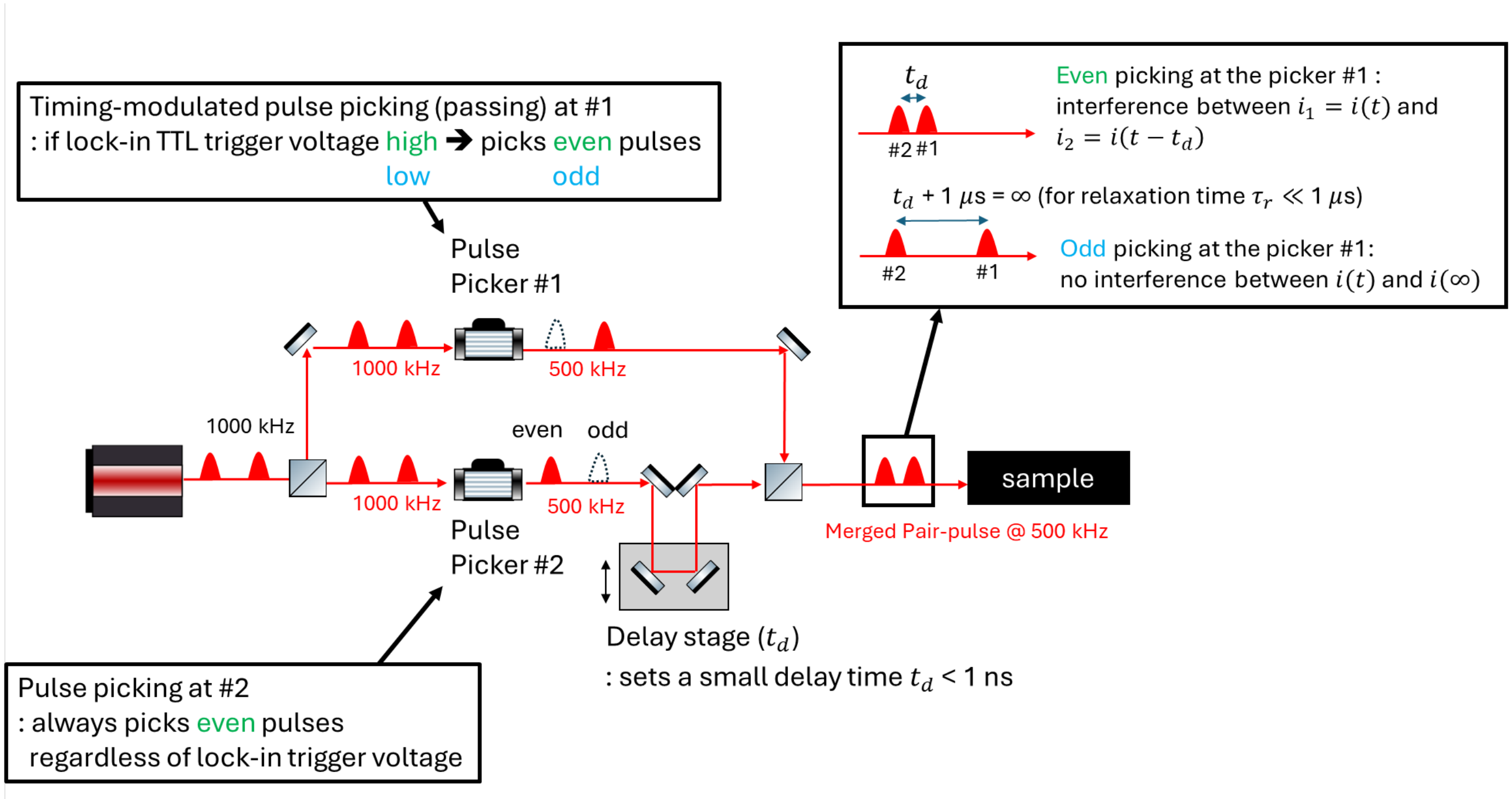

**Supplementary Fig. 3 | Digital delay time modulation for lock-in measurement of pair-pulse correlation.**

Detailed explanation of each component can be found in Methods Sec. 4.

**Supplementary Fig. 4**

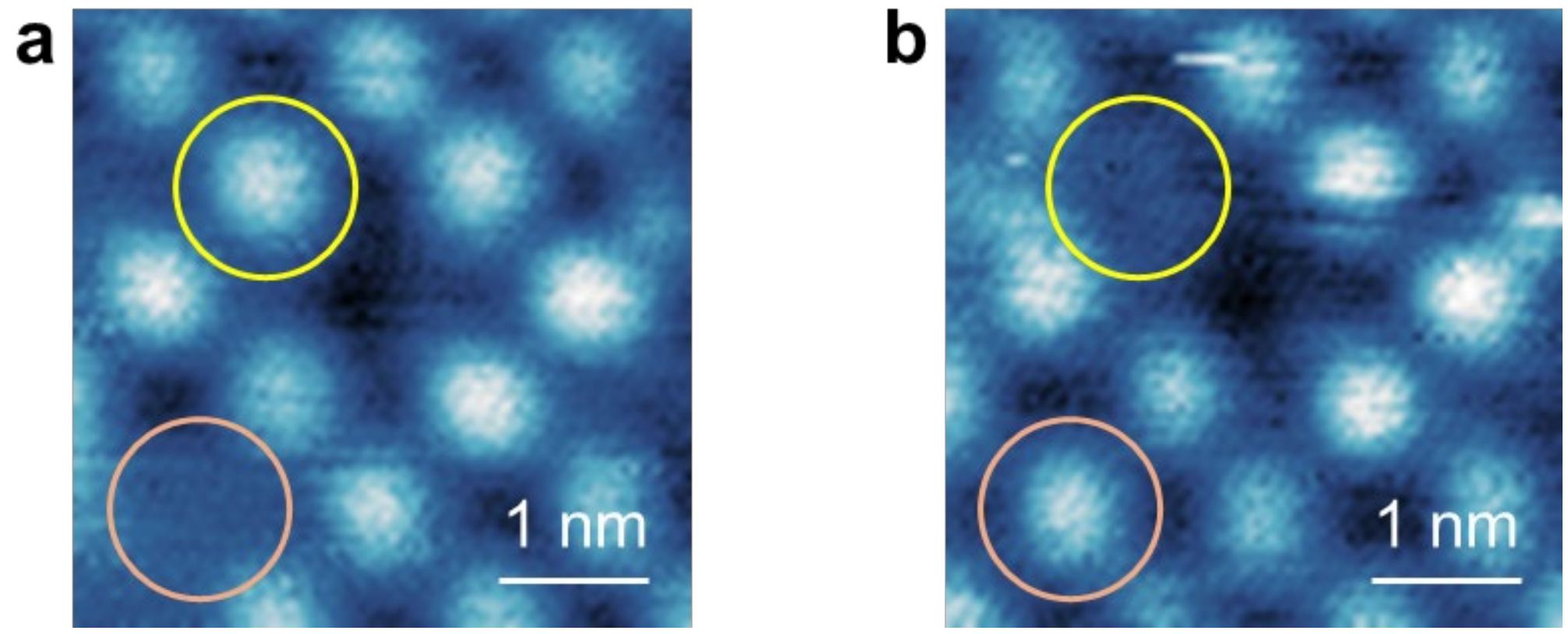

**Supplementary Fig. 4 | Surface iodine atom motion under laser illumination.**

**a,** Topography of $(TaSe_4)_2I$ under laser illumination ($V_{bias}$ = -2.00 V, I = 2.50 pA); one of the iodine atoms is marked in yellow, while a vacancy is marked in orange. **b,** Another topography after ~2 mins. of laser illumination (also at $V_{bias}$ = -2.00 V, I = 2.50 pA) shows that the iodine atom marked in **a** has jumped into the vacancy from **a**. All data shown is taken at T = 77 K with a fluence of 0.265 $mJ/cm^2$.

**Supplementary Fig. 5**

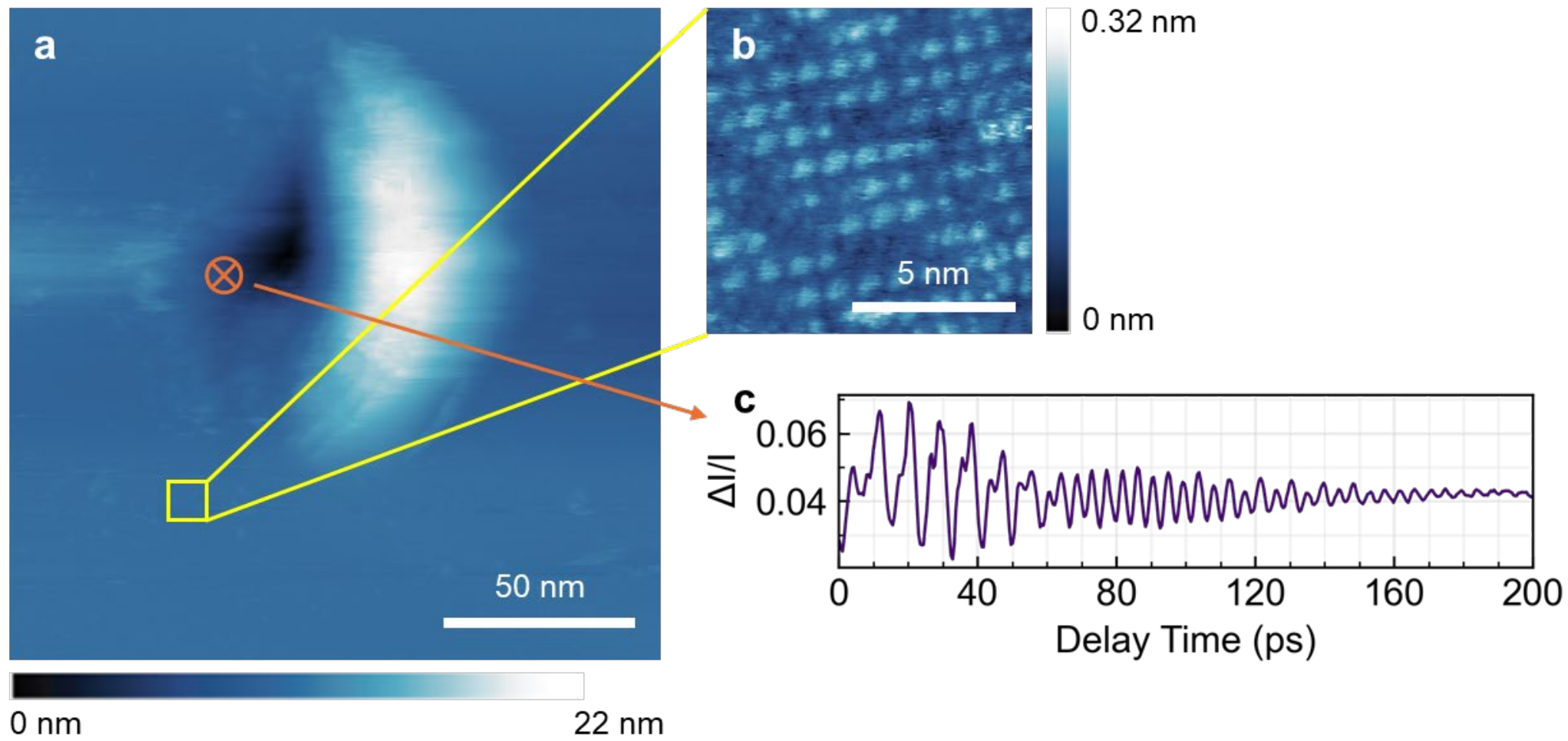

**Supplementary Fig. 5 | STM topographies near point-contact region and an example UPC delay-time sweep.**

**a,** STM topography in a region where a set of UPC measurements were taken. The tip position was extended towards the sample from tunneling position during this set of measurements; the position where the tip made a contact to the sample is marked with an orange cross. Topography parameters are $V_{bias}$ = -1.00 V and $I$ = 4.00 pA. **b,** Topography within 100 nm of the point-contact position after the UPC measurements shows atomically resolved lattice; $V_{bias}$ = -1.00 V and $I$ = 5.00 pA. **c,** An example UPC delay-time sweep at the spot marked in (a); $V_{bias}$ = -10 mV, $I$ = 26.3 pA, $FL$ = 0.145 mJ/cm$^2$ for each pump during this sweep. Data shown in this figure were taken at a temperature of 77 K.

## Supplementary Fig. 6

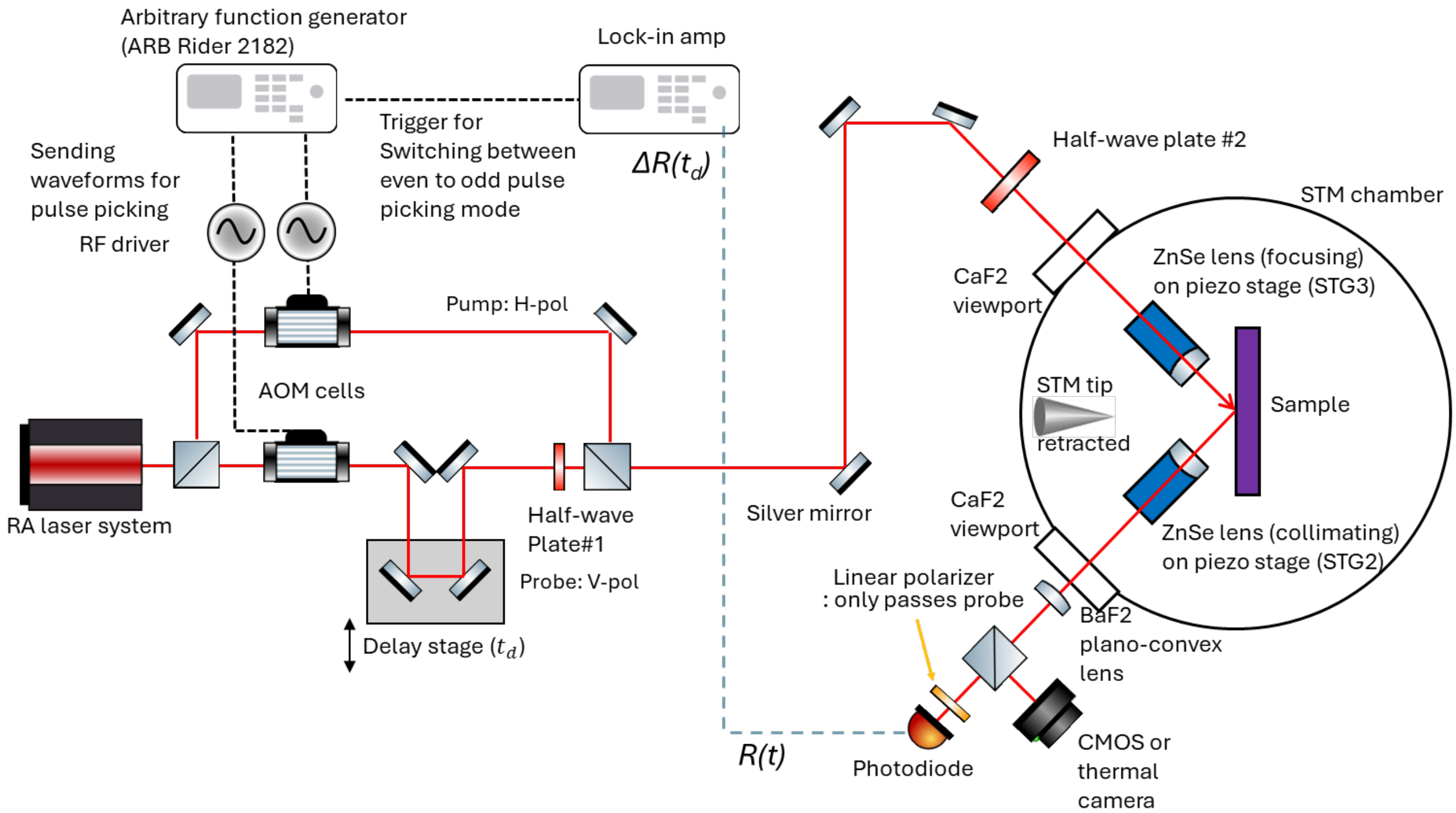

**Supplementary Fig. 6 | Optics layout for the optical pump-probe reflectance (OPPR) with pair-pulse correlation scheme.**

Detailed explanation of the measurement procedure can be found in Methods Sec. 4-5.

**Supplementary Fig. 7**

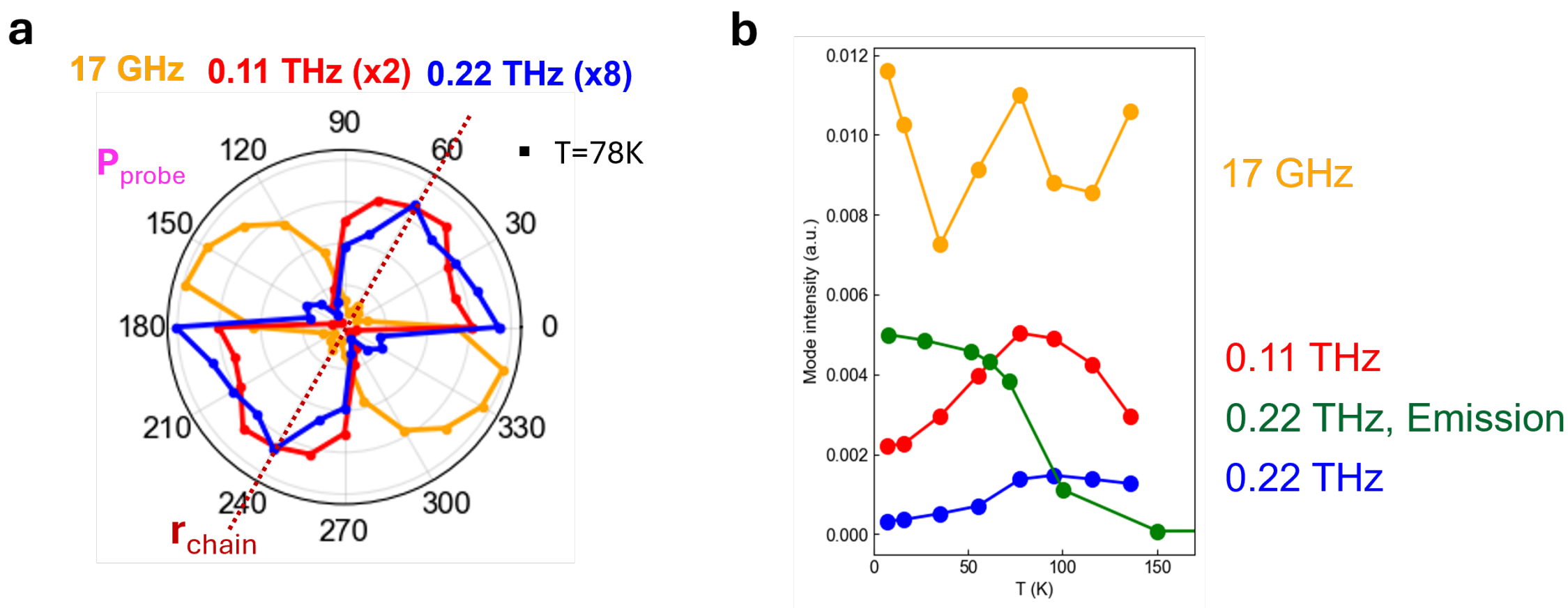

**Supplementary Fig. 7 | Probe polarization and temperature dependence of the OPPR modes.**

**a,** The mode intensities of the OPPR modes with respect to the probe polarization $\mathbf{P}_{probe}$. The measurement was taken at 77 K. During the polarization sweep, pump fluence was fixed to 0.174 mJ/cm$^2$. The probe fluence ranged from 0.426 (when $\mathbf{P}_{probe}$ = 0 deg) to 1.236 (90 deg) mJ/cm$^2$ to fix the in-plane component of the laser electric field from the probe to 1.15 MV/cm. The red dotted line with $\mathbf{r}_{chain}$ denotes the direction of Ta-atom-chain (*c*-axis of the crystal). The mode intensities of 0.11 and 0.22 THz modes are multiplied for clarity. The 17 GHz mode is polarized perpendicular to the chain direction whereas the 0.11 and 0.22 THz modes are polarized along the chain direction. **b**, The temperature dependence of all the OPPR modes. The measurement conditions are the same as Fig. 4d in the main text.

**Supplementary Fig. 8**

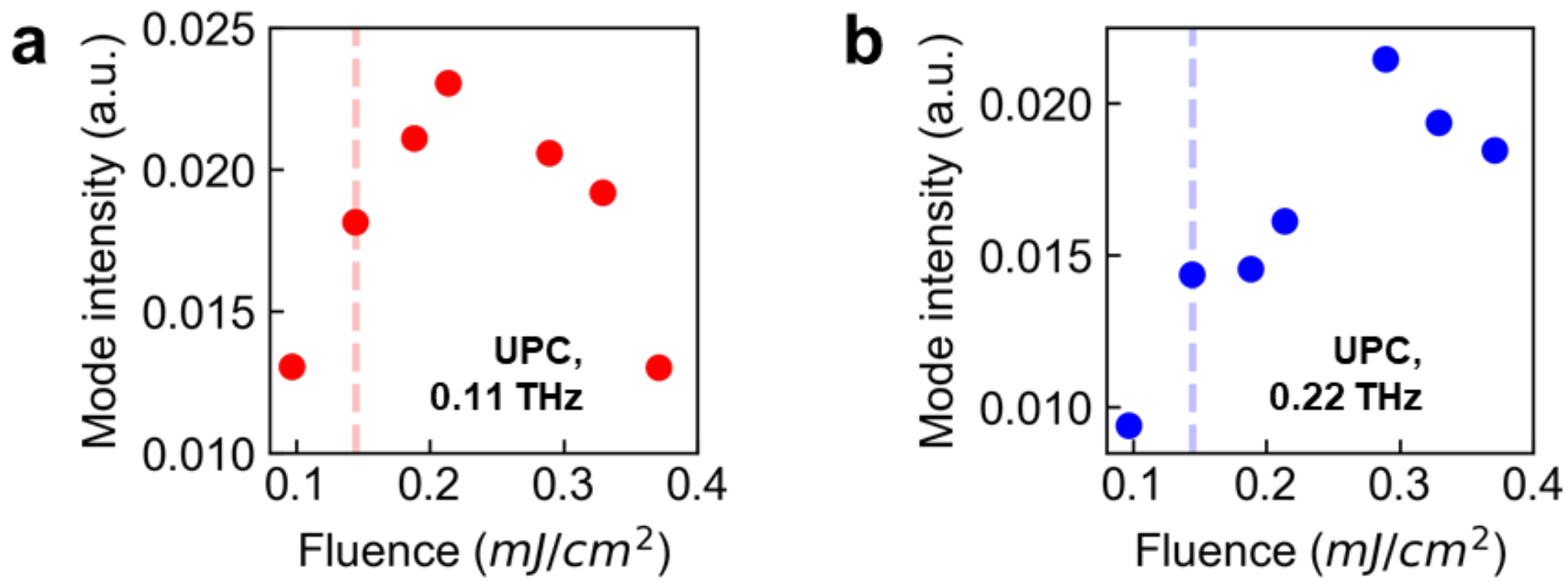

**Supplementary Fig. 8 | Fluence dependence of UPC modes.**

Fluence dependence of the (**a**) 0.11 THz and (**b**) 0.22 THz modes in UPC at $T$ = 77 K, $Z_{ext}$ = 25 nm. The dashed line marks the fluence of 0.145 mJ/cm$^2$, which is the fluence used for UPC measurements presented in the main text. Both UPC modes show linear fluence dependence at low fluence regime (< 0.2 mJ/cm$^2$). The fact that 0.22 THz mode shows linear dependence would imply it is not a 2$^{nd}$ harmonic mode of 0.11 THz mode. At high fluence regime, both modes show a downturn. This could be due to heating effects or increase in thermally excited quasiparticle density, which could increase scattering of UPC modes and screens long-range Coulomb interactions.

**Supplementary Fig. 9**

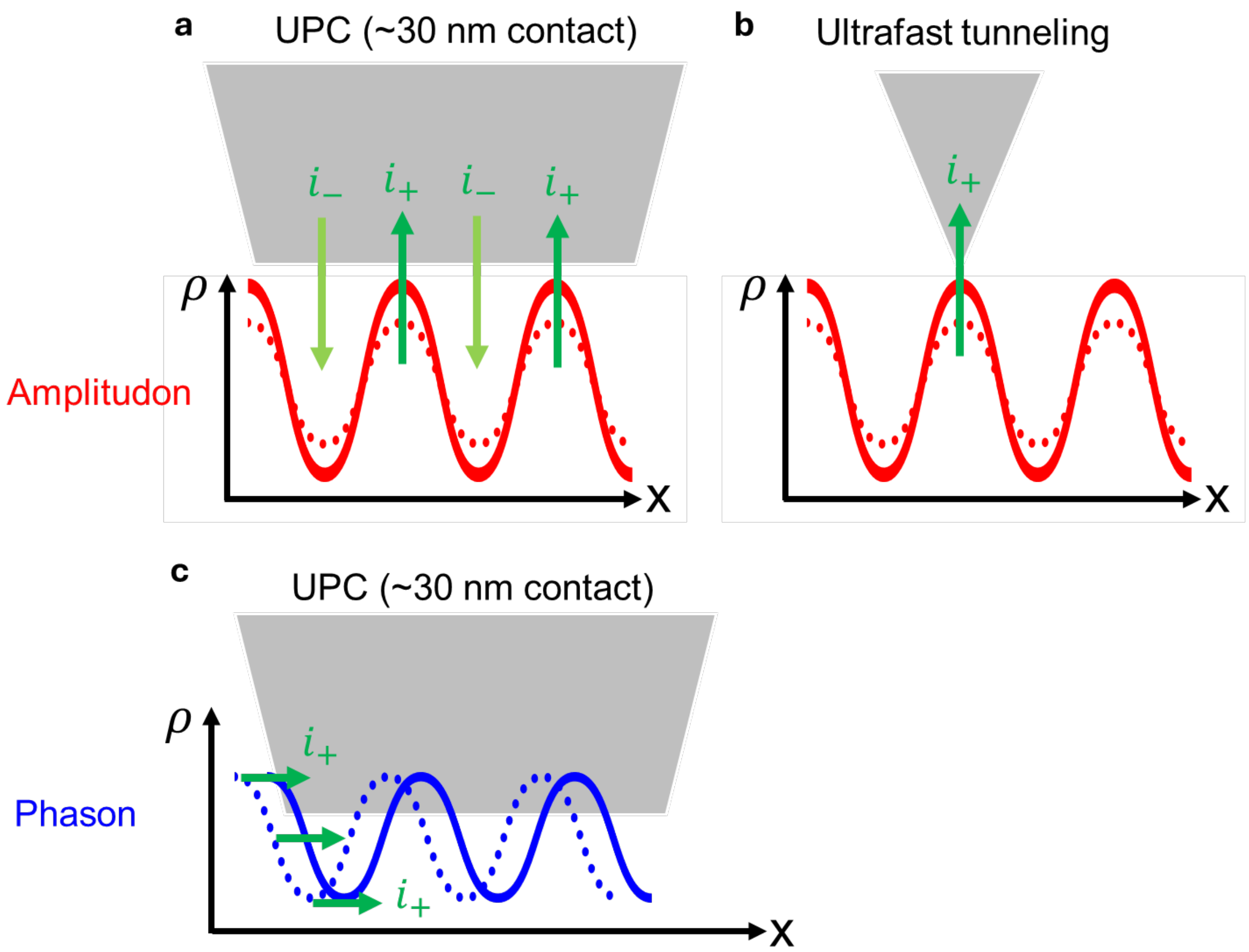

**Supplementary Fig. 9 | Sensitivity of the current oscillation from amplitudon in UPC and ultrafast tunneling.**

**a,** A sketch of motion of charge density $\rho$(x) during CDW amplitude oscillation (red solid and dotted lines). In the case of UPC whose contact laterally spans ≈ 30 nm that is > 3× CDW wavelength of $(TaSe_4)_2I$ (≈ 9 nm), the peak and valley positions within a CDW wavelength register currents at opposite sign to the contact ($i_+, i_-$), resulting in cancellation with each other. **b**, For the case of ultrafast tunneling such as in the case of THz-STM, such cancellation does not occur as its lateral probing size is much smaller than the CDW wavelength. **c**, A sketch of motion of $\rho$ during CDW phase oscillation (blue solid and dotted line) with green arrows indicating the direction of motions at the peak and valley positions. In the case of phase sliding mode, the entire wave moves in the same direction, registering currents that add up constructively, resulting in a measurable signal in UPC.

**Supplementary Fig. 10**

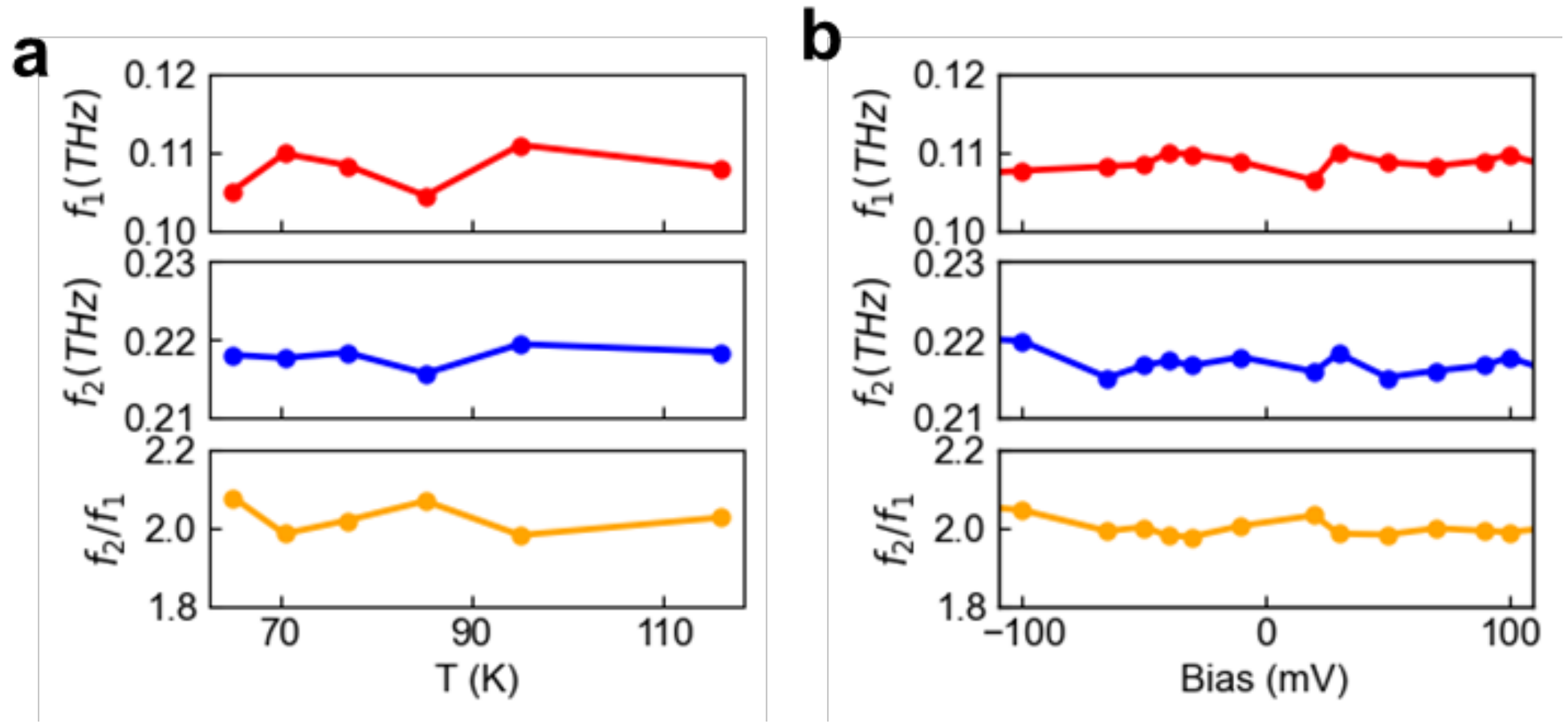

**Supplementary Fig. 10 | UPC Mode frequencies and their ratio over various temperatures and biases.**

**a**, The mode frequencies of the 0.11 THz ($f_1$,red), 0.22 THz ($f_2$,blue) UPC modes. Their ratio $f_2/f_1$ (orange) shows nearly constant value ≈2. **b**, The UPC mode frequencies $f_1$, $f_2$, and their ratio $f_2/f_1$, showing nearly a constant ratio ≈2 over the biases within the gap.

**Supplementary Fig. 11**

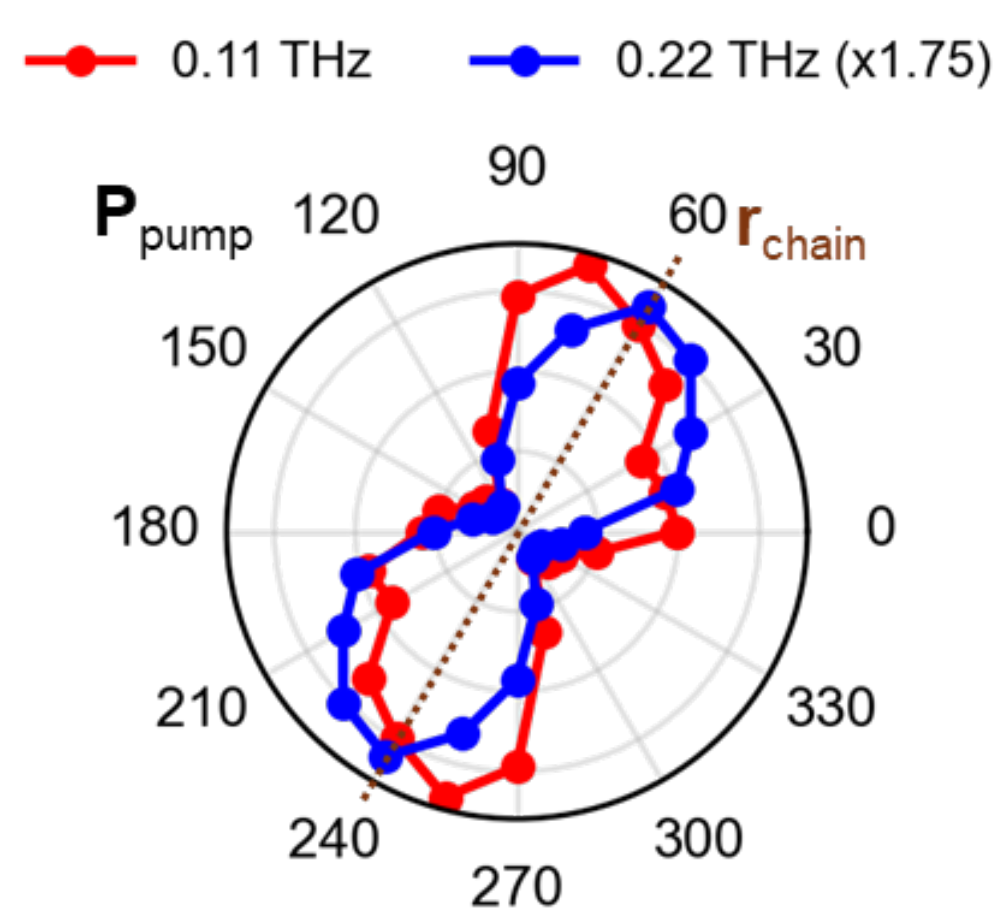

**Supplementary Fig. 11 | Pump polarization dependence of the UPC mode intensities.**

Pump polarization $\mathbf{P}_{pump}$ dependence of 0.11 THz (red) and 0.22 THz (blue) mode intensities in UPC measurement at 85 K as an example. $V_{bias}$ = - 10 mV, $Z_{ext}$ = 25 nm. The dotted brown line denotes the c-axis chain direction of the crystal ($\mathbf{r}_{chain}$). Both mode intensities show a two-fold pattern whose peaks are centered near $\mathbf{r}_{chain}$. In-plane electric field of each pump is fixed at 0.56 MV/cm for this polarization sweep. The intensity of 0.22 THz mode is multiplied for clarity.

**Supplementary Fig. 12**

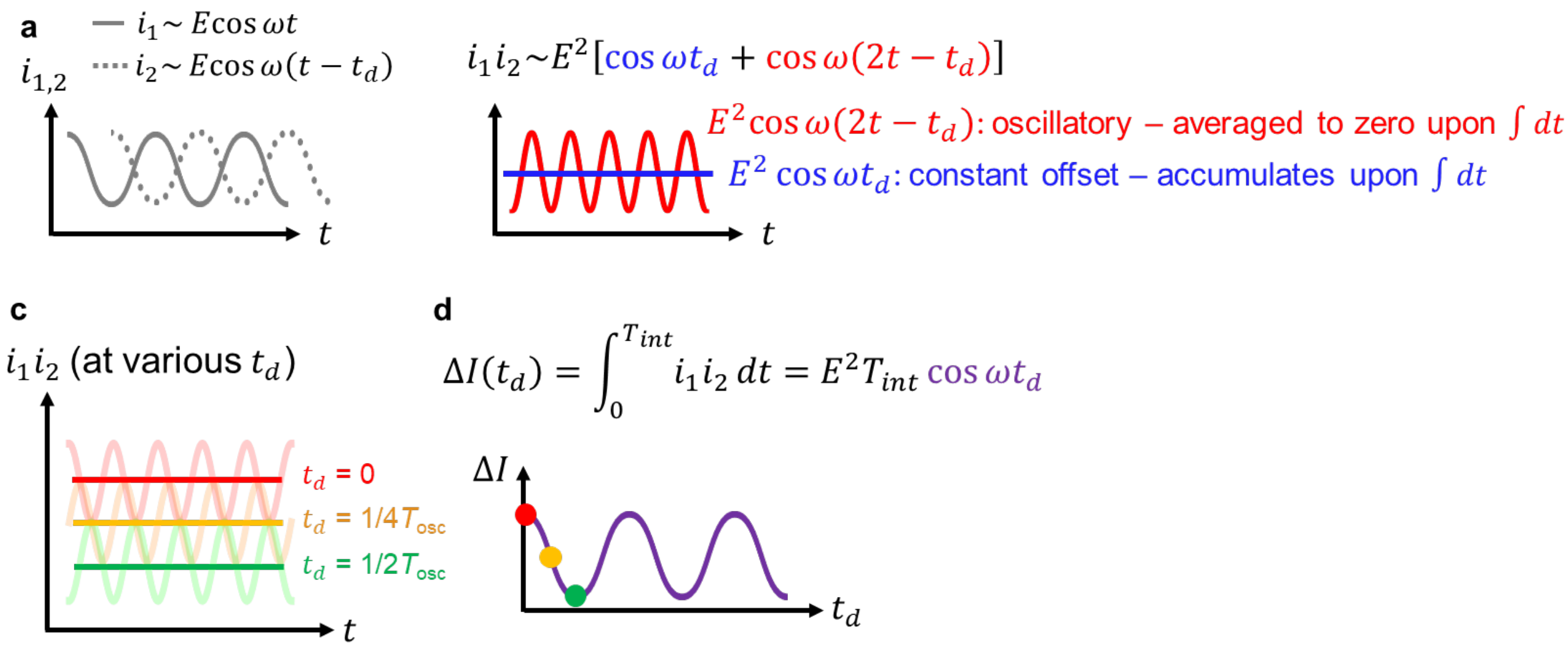

**Supplementary Fig. 12 | Illustration of how the transient current $\Delta I(t_d)$ (time-averaged correlation current) captures collective modes of the system.**

**a**, Ultrafast currents $i_1(t)$, $i_2(t)$ in terms of real time $t$ due to collective modes excited by pump #1,#2. **b**, Correlation current $i_1 i_2(t, t_d)$ plotted in terms of $t$. It consists of constant offset as a function of $t_d$ (blue line) and oscillatory term with respect to $t$ (red line). Upon integration over time $t$ ($\int i_1 i_2 dt$), the oscillatory term is averaged to nearly zero while the constant term is accumulated, and hence $\int i_1 i_2 dt$ is dominated by the offset. **c**, Illustration of $i_1 i_2(t)$ for various values of $t_d$. The thick solid line denotes the constant offset which varies with $t_d$ (~ $\cos\omega t_d$). Here, $T_{osc} = 2\pi/\omega$ is the oscillation period of $i(t)$. **d**, The transient current $\Delta I(t_d)$, which captures time-averaged correlation current, oscillates in delay time $t_d$ at the same frequency $\omega$ as the ultrafast current $i_{1,2}(t)$ from the collective modes.

**Supplementary Fig. 13**

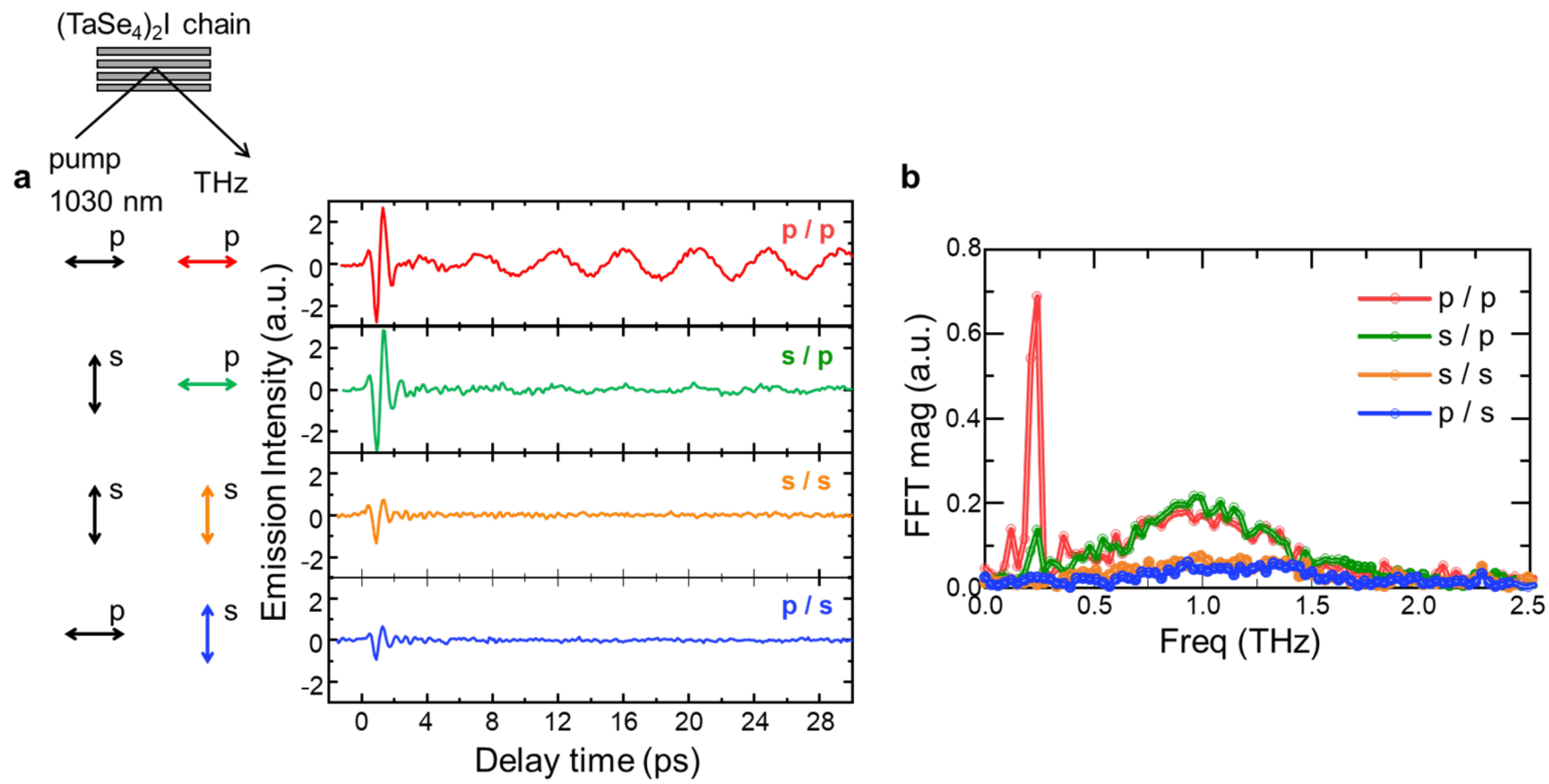

**Supplementary Fig. 13 | Pump polarization dependence of the optical-pump THz emission 0.22 THz mode intensity.**

Optical-pump (1030 nm) THz-emission intensities with oblique incidence of pump (45 degrees from the surface normal) at T = 7 K and fluence = 0.58 mJ/cm$^2$. The chain direction is along the *c*-axis. **a**, Time-domain emitted THz signal. The left column denotes the polarization direction of the pump. The right column denotes the polarization component of the emitted THz signal. **b**, Frequency-domain emitted THz signal. The 0.22 THz mode is prominent with p-polarized pump (red curve), but much suppressed with s-polarized pump (green curve).